\documentclass[12pt]{article}
\usepackage[utf8]{inputenc}
\usepackage{amsmath,amssymb,amsfonts,graphics,graphicx,amscd,amsfonts,epsfig,color}
\usepackage{multirow}
\usepackage{subfig}
\usepackage{cite}
\usepackage{enumitem}
\usepackage{here}

\newcommand {\beq} {\begin{equation}}
\newcommand {\eeq} {\end{equation}}
\newcommand {\beqa}{\begin{eqnarray}}
\newcommand {\eeqa}{\end{eqnarray}}

\newcommand {\tr}{{\textrm{tr}\,}}
\newcommand {\Tr}{\mbox{Tr\,}}

\newcommand {\Pf}{\mbox{Pf}}

\newcommand{\rf}       [1]{(\ref{#1})}
\newcommand{\vev}      [1]{{\ensuremath {\langle {#1} \rangle}}}
\newcommand{\mf}          {\ensuremath{m_{\rm f}}}
\newcommand{\epsilonm}    {\ensuremath{\epsilon}}
\newcommand{\Sb}          {\ensuremath{S_{\mathrm{b}}}} 
\newcommand{\Sf}          {\ensuremath{S_{\mathrm{f}}}} 
\newcommand{\Seff}        {\ensuremath{S_{\mathrm{eff}}}}

\makeatletter
    
    \@addtoreset{equation}{section}
  \makeatother

\allowdisplaybreaks[4]

\date{}
\begin{document}

\begin{titlepage}
\renewcommand{\thefootnote}{\fnsymbol{footnote}}

\begin{flushright} 
KEK-TH-2184
\end{flushright} 

\vspace{0.1cm}

\begin{center}
{\bf \large Complex Langevin analysis of 
the spontaneous breaking of 10D rotational symmetry
in the Euclidean IKKT matrix model}
\end{center}

\vspace{0.1cm}
\vspace{0.1cm}

\begin{center}

         Konstantinos N. A{\sc nagnostopoulos}$^{1)}$\footnote
          { E-mail address : konstant@mail.ntua.gr},  
 
         Takehiro A{\sc zuma}$^{2)}$\footnote
          { E-mail address : azuma@mpg.setsunan.ac.jp},  
         Yuta I{\sc to}$^{3,4)}$\footnote
          { E-mail address : y-itou@tokuyama.ac.jp},
         Jun N{\sc ishimura}$^{4,5)}$\footnote
          { E-mail address : jnishi@post.kek.jp},

         Toshiyuki O{\sc kubo}$^{6)}$\footnote
          { E-mail address : tokubo@meijo-u.ac.jp} and 
          
         Stratos Kovalkov P{\sc apadoudis}$^{1)}$\footnote
          { E-mail address : sp10018@central.ntua.gr}         
 
\vspace{0.5cm}
$^{1)}$ {\it Physics Department, National Technical University,\\
Zografou Campus, GR-15780 Athens, Greece}

$^{2)}${\it Institute for Fundamental Sciences, Setsunan University, \\
17-8 Ikeda Nakamachi, Neyagawa, Osaka, 572-8508, Japan}

$^{3)}${\it National Institute of Technology, Tokuyama College,\\
 Gakuendai, Shunan, Yamaguchi 745-8585, Japan}

$^{4)}${\it KEK Theory Center, 
High Energy Accelerator Research Organization,\\
1-1 Oho, Tsukuba, Ibaraki 305-0801, Japan} 

$^{5)}${\it Graduate University for Advanced Studies (SOKENDAI),\\
1-1 Oho, Tsukuba, Ibaraki 305-0801, Japan} 

$^{6)}${\it Faculty of Science and Technology, Meijo University,\\
Nagoya, 468-8502, Japan} 

\end{center}

\vspace{0.5cm}


\begin{abstract}
\noindent
The IKKT matrix model 
is a promising candidate for
a nonperturbative formulation of superstring theory,
in which spacetime is conjectured to emerge dynamically 
from the microscopic matrix degrees of freedom in
the large-$N$ limit.
Indeed in the Lorentzian version, Monte Carlo studies
suggested the emergence of (3+1)-dimensional expanding space-time.
Here we study the Euclidean version instead,
and investigate an alternative scenario for dynamical
compactification of extra dimensions via the spontaneous symmetry breaking
(SSB) of 
10D rotational symmetry.
We perform numerical
simulations based on the complex Langevin method (CLM) 
in order to avoid 
a severe sign problem. 
Furthermore, in order to avoid the singular-drift problem in the CLM,
we deform the model and determine the SSB
pattern as we vary the deformation parameter. 
From these results,
we conclude that
the original model has an SO($3$) symmetric vacuum,
which is consistent with previous results obtained by the
Gaussian expansion method (GEM).
We also apply the GEM to the
deformed matrix model and find consistency
with the results obtained by the CLM.

\end{abstract}
\vfill
\end{titlepage}
\vfil\eject


\setcounter{footnote}{0}

\section{Introduction}

Superstring theory has been studied intensively
as a unified theory that includes quantum gravity. 
The theory is defined in ten spacetime dimensions and 
the connection to the real world, where only four
dimensions are macroscopic, is realized via compactification of
the extra dimensions. How this can actually occur
has been investigated
perturbatively by using D-brane configurations as a background, 
leading to tremendously many vacua giving rise to
the so-called string landscape. 
Clearly, it is important to see if this picture remains
valid
when the issue is addressed in a fully nonperturbative manner.

The IKKT matrix model \cite{Ishibashi:1996xs}, also known as the
type IIB matrix model, was proposed as a nonperturbative formulation of
superstring theory.  Formally, the action of the model can be obtained
by dimensionally reducing the action of 10D $\mathcal{N}=1$ SU($N$) super
Yang-Mills (SYM) theory to 0D. In this model, spacetime emerges
dynamically from the eigenvalues of the ten bosonic 
matrices
in the large-$N$ limit \cite{Aoki:1998vn},
which enables
the scenario for dynamical compactification 
as a purely nonperturbative effect.

Evidence supporting
such a scenario
has been provided by
simulating the Lorentzian version of the IKKT matrix 
model \cite{Kim:2011cr,Ito:2013qga,Ito:2013ywa,Ito:2015mxa,Ito:2015mem,Ito:2017rcr,Azuma:2017dcb,Aoki:2019tby,Nishimura:2019qal}.  
In these simulations, 
continuous time
emerges dynamically and three-dimensional space undergoes expansion after
a critical time, with the six extra dimensions remaining small.
This is highly nontrivial since time is given by 
the ordered eigenvalues of the temporal matrix $A_0$,
and in the SU($N$) basis used to diagonalize $A_0$, the
dominant configurations of the spatial matrices $A_i$ have a band diagonal 
structure, from which 
one can read off the time evolution of space.
By taking the large-$N$ limit,
the eigenvalue distribution of $A_0$
extends in physical units
and a sensible continuum limit can be defined, a fact that emerges
also from the dynamics of the model.
The results in Ref.~\cite{Ito:2013ywa} suggest that the expansion 
is exponential at early times,
which turns into a power law at later times \cite{Ito:2015mxa}, 
providing evidence that a realistic cosmological scenario 
may also arise dynamically from this model.

Monte Carlo simulations of the Lorentzian model
are hindered by a severe sign problem 
coming from the phase factor ${\rm e}^{i \Sb}$, 
where $\Sb$ is the bosonic part of the
action. In the early work
\cite{Kim:2011cr,Ito:2017rcr,Ito:2013ywa,Ito:2013qga,Ito:2015mem,Ito:2015mxa,Azuma:2017dcb,Aoki:2019tby},
this problem was avoided by first integrating out the scale factor of the
bosonic matrices, yielding a function of $\Sb$ which is 
sharply peaked near the origin. Then, by approximating this function by
a sharply peaked Gaussian function, the sign problem was avoided.
This, however, leads to singular spatial configurations, 
showing that it is highly nontrivial 
to obtain 3D expanding space with 
a smooth structure \cite{Aoki:2019tby}.
Recent work, which avoids this approximation but confronts 
the sign problem, 
provided evidence that the 3D expanding space can 
have a smooth structure in the large-$N$ limit \cite{Nishimura:2019qal}. 
This work also shows that 
the complex Langevin method (CLM) \cite{Parisi:1984cs,Klauder:1983sp} 
can be used successfully in circumventing the sign problem 
in this model.

The Euclidean version of the IKKT matrix model 
and related models, on the other hand,
have been studied numerically
\cite{Hotta:1998en,Ambjorn:2000bf,Ambjorn:2000dx,Ambjorn:2001xs,Anagnostopoulos:2001yb,Anagnostopoulos:2013xga,Anagnostopoulos:2015gua,Anagnostopoulos:2017gos}
since long time
before numerical studies of the Lorentzian version were started.
This is because
it can be thought of as being the direct analog of lattice QCD 
for superstring theory,
and moreover it was shown \cite{Krauth:1998xh,Austing:2001pk}
to have a finite partition function without introducing infrared cutoffs 
unlike the Lorentzian version. 
However, the simulations are hindered by a severe sign problem 
here as well because
the Pfaffian obtained after integrating out the fermionic matrices
is complex in general.

In the Euclidean model, the scenario for dynamical compactification of extra
dimensions is expected to be realized via spontaneous symmetry breaking
(SSB) of the SO($10$) rotational symmetry
due to the wild fluctuations of the phase of the Pfaffian
for SO($d$) symmetric configurations
with larger $d$ \cite{Nishimura:2000ds,Nishimura:2000wf}.
A strong evidence for the SSB
has been provided by studying the model using the
Gaussian expansion method (GEM).
This method was applied to the Euclidean IKKT model
and related matrix models
\cite{Nishimura:2001sx,Kawai:2002jk,Aoyama:2006rk,Aoyama:2010ry,Nishimura:2011xy}
realizing SSB to lower dimensional spaces.
In particular, it was shown \cite{Nishimura:2011xy} 
in the IKKT matrix model 
that the SO($3$) symmetric vacuum has the lowest free energy, 
which implies SSB to SO($3$). The ratio between the
three extended directions and the seven shrunken directions was also
calculated and it was found to be finite. 
This SSB can be naturally attributed to the effect of the phase
of the Pfaffian since the phase-quenched simulations of the
model showed no SSB \cite{Ambjorn:2000dx}.
A direct confirmation of this scenario
by numerical simulation, however, 
requires 
some new ideas to overcome the sign problem.
In a series of 
work\cite{Anagnostopoulos:2001yb,Anagnostopoulos:2013xga,Anagnostopoulos:2015gua},
the Euclidean version of the IKKT matrix model
was studied
by using a density of states based method
\cite{Anagnostopoulos:2001yb,Ambjorn:2002pz,Ambjorn:2004jk,Anagnostopoulos:2010ux,Anagnostopoulos:2011cn}, which was successful in that
it made it possible 
to obtain the extent of space in each direction
for a given pattern of SSB although
it was not powerful enough to determine the SSB pattern itself.

In fact, the sign problem 
occurs
in Monte Carlo simulation of various interesting
systems such as finite density QCD, supersymmetric theories, 
strongly correlated electron systems and real-time quantum field theories.
If one uses the reweighting method to simulate the system,
the computational cost increases exponentially with the system size.
In recent years there has been major progress in
evading the sign problem by complexifying the dynamical degrees of
freedom of the system under study. 
One of such methods is the
generalized Lefschetz-thimble method
\cite{Cristoforetti:2012su,Cristoforetti:2013wha,Fujii:2013sra,Alexandru:2015sua,Fukuma:2017fjq,Fukuma:2019uot}, which amounts to deforming the integration contour in such a way that
the sign problem becomes mild enough to be handled by the reweighting
method. Another approach is the CLM
\cite{Parisi:1984cs,Klauder:1983sp}, which extends the idea
of stochastic quantization \cite{Parisi:1980ys}
and defines a
stochastic process for the complexified variables so that the
expectation values with respect to this process are equal to the
expectation values defined in the original system. The use of the CLM allows
one to study large systems, but it has the problem of not 
always yielding correct results. (See, for instance, 
Refs.~\cite{Ambjorn:1985cv,Ambjorn:1986mf} for early work.)
Recently, the conditions 
for the correct convergence
were clarified 
\cite{Aarts:2009uq,Aarts:2011ax,Nishimura:2015pba,Nagata:2016vkn,Salcedo:2016kyy,Nagata:2018net}
and various new techniques have been proposed
to meet these conditions
for a large space of parameters\cite{Seiler:2012wz,Nagata:2015uga,Tsutsui:2015tua,Ito:2016efb,Nagata:2016alq,Doi:2017gmk}.
Thanks to these developments,
the CLM has been applied successfully to many systems in lattice quantum
field theory 
\cite{Berges:2005yt,Berges:2006xc,Berges:2007nr,Pehlevan:2007eq,Aarts:2008rr,Aarts:2008wh,Aarts:2009hn,Sexty:2013ica,Fodor:2015doa,Aarts:2016qrv,Attanasio:2018rtq,Nagata:2018net,Nagata:2018mkb,Ito:2018jpo,Kogut:2019qmi,Sexty:2019vqx,Tsutsui:2019suq}
and matrix models 
\cite{Mollgaard:2013qra,Mollgaard:2014mga,Ito:2016efb,Bloch:2017sex,Anagnostopoulos:2017gos,Nishimura:2019qal,Nagata:2018net,Basu:2018dtm,Joseph:2019sof}
with complex actions. (For a recent review on the CLM and related methods, 
see Ref.~\cite{Berger:2019odf}.)

In Ref.~\cite{Anagnostopoulos:2017gos},
we applied the CLM 
to the 6D version of the Euclidean IKKT matrix model,
and obtained results consistent with 
the ones obtained by the GEM \cite{Aoyama:2010ry}, i.e., 
SSB to SO($3$).
We used a technique \cite{Ito:2016efb}
to avoid the singular-drift problem \cite{Nishimura:2015pba}
caused by the eigenvalues of the Dirac operator that accumulate 
near zero \cite{Mollgaard:2013qra}.
The model is deformed by introducing a parameter $\mf$,
which corresponds to adding a mass-like term in the Dirac operator.
Thus the condition \cite{Nagata:2016vkn}
that ensures the correctness of the CLM can be met, and
one can make an extrapolation $\mf \rightarrow 0$ 
to obtain the results 
for the original model. 

In this paper we apply the same method to the original 
Euclidean IKKT matrix model. 
The simulations are more challenging than those of the 6D version
because the number of the fermionic matrices
increases by a factor of four.
We find that SSB to lower dimensional space
occurs as the deformation
parameter $\mf$ is reduced and that our data
are consistent with an SO($3$) symmetric vacuum appearing 
at $\mf = 0$
as predicted by the GEM for the undeformed model \cite{Nishimura:2011xy}. 
These results are in sharp contrast to the ones obtained
in Ref.~\cite{Anagnostopoulos:2015gua} using
the density of states based method, where calculations had to be
performed for each vacuum with SO($d$) symmetry.
We also find at larger $N$
that
the singular-drift problem becomes milder,
which enables the use of a smaller 
deformation parameter in the simulations. 
This gives us a hope that a more complete understanding of 
the Euclidean IKKT matrix model may be possible by
extending this work to larger $N$.
%



Our calculations may still suffer from finite-$N$ effects
and the results can be sensitive to systematic errors 
introduced by extrapolations including that for large $N$.
We therefore perform a consistency check 
by applying the GEM to the deformed IKKT matrix model.
We calculate the free energy
up to three loops
and obtain physical solutions 
for the self-consistency equations
with the SO($d$) ($d=6,7$) ansatzes.
We find that the SO(6) symmetric vacuum has smaller free
energy as $\mf$ is decreased.
We also calculate
the extent of space 
and find consistency with the results
obtained by the CLM.

The rest of this paper is organized as follows. 
In Section \ref{sec_IKKT}, we
give a brief review of the Lorentzian and Euclidean
versions of the IKKT matrix model. In Section \ref{sec:mc}, we
apply the CLM to the Euclidean IKKT matrix
model with a mass deformation of the fermionic action
and present our results. In Section \ref{sec_GEM}, we 
apply the GEM to the deformed Euclidean IKKT matrix model
and compare the results with those of the CLM. 
Section \ref{sec:summary} is devoted to
a summary and discussions.
In Appendix \ref{sec:large-N-singular-drift},
we show some results suggesting that the singular-drift problem
vanishes at large $N$.

\section{Brief review of the IKKT matrix model} \label{sec_IKKT}

The action of the IKKT matrix model \cite{Ishibashi:1996xs}
is given by
\begin{eqnarray}
 S &=& S_{\textrm{b}} + S_{\textrm{f}} \ , \textrm{ where } 
\label{IKKT_action} \\
 S_{\textrm{b}} &=& - \frac{1}{4} N \, \textrm{tr} 
[A_{\mu}, A_{\nu}] [A^{\mu}, A^{\nu}] \ , 
\label{IKKT_boson} \\
 S_{\textrm{f}} &=& - \frac{1}{2} N \,  \textrm{tr} \left( 
\psi_{\alpha} 
({\cal C} \Gamma^{\mu})_{\alpha \beta} [A_{\mu}, \psi_{\beta}] \right) \ . 
\label{IKKT_fermion}
\end{eqnarray}
The vectors $A_{\mu}$ ($\mu =0,1,2, \ldots, 9$) are $N \times N$
traceless Hermitian matrices, and the Majorana-Weyl spinors
$\psi_{\alpha}$ ($\alpha =1,2, \ldots, 16$) are $N \times N$ traceless
matrices with Grassmann entries. 
The $16 \times 16$ matrices
$\Gamma^{\mu}$ and
${\cal C}$ 
are the 
gamma matrices after Weyl projection
and the charge conjugation matrix, respectively, in ten dimensions.

\subsection{the Lorentzian version}
In this section we review the Lorentzian version of the IKKT matrix
model, in which the indices are contracted using the Minkowski metric
$\displaystyle \eta_{\mu \nu} = \textrm{diag} (-1,1,1,\ldots,1)$.
The model is invariant under SO($9,1$) Lorentz transformations, which
act on the vectors $A_\mu$ and the Majorana-Weyl spinors $\psi_\alpha$.
The model also possesses the SU($N$) symmetry
\begin{equation}
\label{i01}
A_\mu\to U^\dagger A_\mu U \  , \quad \quad
\psi_\alpha\to U^\dagger \psi_\alpha U \  ,
\end{equation}
which is inherited from the gauge invariance 
of the 10D ${\cal N}=1$ SYM action after reduction to 0D. 
The supersymmetry of the SYM theory, on the other hand,
enhances to an ${\cal N}=2$ supersymmetry in the IKKT 
matrix model \cite{Ishibashi:1996xs}.
This allows us to interpret the eigenvalues
of the matrices $A_\mu$ as the $N$ points 
in the target spacetime \cite{Aoki:1998vn}, which 
are expected to represent the continuum spacetime in the
large-$N$ limit\footnote{The matrices $A_\mu$ cannot 
in general be diagonalized simultaneously,
so this can lead to a ``fuzzy'' spacetime. 
Whether classical spacetime emerges or not
is a dynamical question.}.

The partition function is given by \cite{Kim:2011cr}
\begin{equation}
 Z = \int dA\, d\psi\, e^{iS} = 
\int dA\, e^{iS_{\textrm{b}}}\, \textrm{Pf } {\cal M} \  , 
\label{LIKKT_partition}
\end{equation}
where the Pfaffian $\textrm{Pf } {\cal M}$
comes from integrating out the fermionic matrices $\psi_\alpha$. 
The $16 (N^2-1) \times 16(N^2-1)$ anti-symmetric matrix
${\cal M}$ is defined by its action
\begin{equation}
 \psi_{\alpha} \to ({\cal M} \psi)_{\alpha} = 
({\cal C} \Gamma^{\mu})_{\alpha \beta} [A_{\mu}, \psi_{\beta}] 
\label{M_definition}
\end{equation}
on the linear space of traceless complex $N \times N$ matrices.
In fact, it turns out that the Pfaffian $\textrm{Pf } {\cal M}$
is real\footnote{Although
$\textrm{Pf } {\cal M}$ can take negative values, it does not cause
the sign problem in the numerical
simulations of 
Refs.~\cite{Kim:2011cr,Ito:2017rcr,Ito:2013ywa,Ito:2013qga,Ito:2015mem,Ito:2015mxa,Azuma:2017dcb}
since
configurations with $\textrm{Pf } {\cal M}<0$ are 
very rare and one can simulate the system by considering
only $|\textrm{Pf } {\cal M}|$.} in the present Lorentzian model.
The bosonic action $\Sb$ can be written as
\begin{equation}
\label{i02}
\Sb = \frac{1}{4} N \,\textrm{tr}\left( F_{\mu\nu} F^{\mu\nu} \right)
    = \frac{1}{4} N \left\{ - 2 \, \textrm{tr} \left( F_{0 i}^2 \right) +  
\textrm{tr} \left( F_{i j}^2 \right)\right\}\  ,
\end{equation}
where $ F_{\mu\nu}=i \,  [A_\mu,A_\nu]$ are Hermitian matrices 
and $i,j=1,\ldots,9$ are spatial indices. 
Since the partition function (\ref{LIKKT_partition}) is
divergent as it is, one has to introduce cutoffs 
in the temporal and spatial directions \cite{Kim:2011cr}. 

Using Eq.~\rf{i01}, it is possible to choose a gauge 
that diagonalizes $A_0$ as
\begin{equation}
 A_0 = \textrm{diag} (\alpha_1, \alpha_2, \ldots, \alpha_N) \ , 
\quad \textrm{where } \alpha_1 < \alpha_2 < \ldots < \alpha_N \ . 
\label{A0_diag}
\end{equation}
In this gauge, the spatial matrices $A_i$ turn out to 
have a band-diagonal structure
and for an appropriate integer $n$, the $n \times n$
submatrices ${\bar A}_i$
\begin{equation}
 ({\bar A_i})_{IJ} (t_\nu) = (A_i)_{\nu+I,\nu+J} \  , 
\textrm{ where } I,J=1,\ldots,n \  ,
\label{band_subA}
\end{equation}
can effectively represent space at time $t_\nu$ defined by
\begin{equation}
 t_\nu = \frac{1}{n} \sum_{I=1}^{n} \alpha_{\nu+I} \ , \label{band_time}
\end{equation}
where $\nu = 0,\ldots, N-n$ 
\cite{Kim:2011cr,Ito:2017rcr,Ito:2013ywa,Ito:2013qga,Ito:2015mem,Ito:2015mxa,Azuma:2017dcb}. 
Time emerges dynamically and it is a nontrivial
dynamical question whether this leads to a continuum time 
in the large-$N$ limit.
In Ref.~\cite{Ito:2013ywa}, 
it was shown to be possible to take a continuum limit 
in the large-$N$ limit 
such that the ``volume'' $\Delta$ and the 
lattice spacing $\epsilon$ in time can be tuned to 
go to $\infty$ and $0$, respectively, 
keeping the product $\epsilon\,\Delta$ fixed.

Using this definition of time, it was 
found \cite{Kim:2011cr,Ito:2017rcr,Ito:2013ywa,Ito:2013qga,Ito:2015mem,Ito:2015mxa,Azuma:2017dcb}
that there exists a critical time $t_{\rm c}$, 
after which three spatial directions 
undergo rapid expansion, whereas the other directions remain small. 
This happens due to spontaneous breaking of the SO($9$) rotational
symmetry down to SO($3$). 
In order to see this, we define the ``moment of inertia tensor''
\begin{equation}
 T_{ij}(t)  = 
\frac{1}{n} \textrm{tr} ({\bar A}_i (t) {\bar A}_j (t) )\  , \label{t_munu_L}
\end{equation}
where the trace here
is over the $I,J$ indices in (\ref{band_subA}),
and obtain its nine eigenvalues
$\lambda_i (t)$
with the ordering $\lambda_1 (t) > \lambda_2 (t) > \ldots > \lambda_9 (t)$.
When the expectation values 
$\vev{\lambda_{i} (t)}$ for $i=1,2,3$ are equal but larger than
$\vev{\lambda_{i} (t)}$ for $i=4,\ldots,9$ in the large-$N$ limit,
we conclude that SSB to SO($3$) occurs.
Using simplified models that
describe the qualitative behavior of the IKKT model 
at early and late times respectively,
it was shown that
at early times the large eigenvalues grow exponentially
$\vev{\lambda_i}\sim {\rm e}^{\Lambda t}$ with $t$ \cite{Ito:2013ywa},
whereas at late times the expansion turns
into a power law $\vev{\lambda_i}\sim t^{1/2}$ \cite{Ito:2015mxa}. 
This gives us a hope that the IKKT model
has the dynamics that contain a realistic cosmology 
with an early time inflationary expansion and 
a late time FRW expansion in the radiation dominated era.

The structure of space was examined recently
in Ref.~\cite{Aoki:2019tby}, and it was
found to be dominated by rather singular spatial configurations, 
whose (3+1)D expanding 
behavior is due to submatrices that are close to the Pauli matrices.
Namely,
the radial distribution of spacetime points 
is such
that two points are located very far, 
whereas the rest accumulate near the origin. 
The reason for the 
domination of such configurations was attributed 
to an approximation used in order to avoid
the sign problem in the simulation. As a result of this approximation, 
one effectively simulates a model (\ref{LIKKT_partition})
with ${\rm e}^{i\Sb}$ replaced by ${\rm e}^{\Sb}$.

Having the factor ${\rm e}^{\Sb}$
in the partition function makes configurations
with the Pauli-matrix structure dominant. 
This can be understood by looking at Eq.~\rf{i02}.
Note that the first term favors configurations such that the $A_i$ commute 
with the $A_0$, whereas the second term 
favors configurations such that the $A_i$ are maximally noncommuting.
The balance of these two terms gives 
the band-diagonal structure important in defining \rf{band_subA}.
In Ref.~\cite{Aoki:2019tby} it was shown that 
configurations with 
the Pauli-matrix structure maximize the second term,
subject to the constraint $\frac{1}{N} \tr (A_i)^2=1$ coming 
from the spatial cutoff introduced in the model to 
make the partition function 
finite \cite{Kim:2011cr}.


The approximation, however, may miss
the important contribution of the configurations
that extremize $\Sb$ instead of maximizing it.
This is suggested indeed in Ref.~\cite{Nishimura:2019qal},
where the $D=6$ bosonic IKKT matrix model was studied numerically
using the CLM in order to avoid the sign problem 
without approximations.
The model was generalized by two parameters $s$ and $k$, 
which correspond to Wick rotations on the worldsheet 
and on the target space, respectively.
The Lorentzian model corresponds to $(s,k)=(0,0)$ and 
the Euclidean model studied in this paper corresponds to
$(s,k)=(1,1)$. 
The results for $(s,k)=(-1,0)$, which correspond
to replacing ${\rm e}^{i\Sb}$ by ${\rm e}^{\Sb}$,
are consistent with the results obtained by
the previous simulations\cite{Kim:2011cr,Ito:2017rcr,Ito:2013ywa,Ito:2013qga,Ito:2015mem,Ito:2015mxa,Azuma:2017dcb} of the Lorentzian model
using the approximation,
and show that the dominant configurations are singular
in that the 3D expanding space has the Pauli-matrix structure. 
The generalized model was also studied in the vicinity of $s = 0$ 
and it was shown for a range of parameters $(s,k)$ that it exhibits 
a (3+1)D expanding behavior,
while the dominant configurations depart from the Pauli-matrix structure 
and the spacetime points are distributed
more smoothly than for $(s,k)=(-1,0)$.

Since an infinite number of (3+1)D expanding classical solutions 
without the Pauli-matrix
structure are known to exist 
\cite{Kim:2011ts,Kim:2012mw,Hatakeyama:2019jyw,Klinkhamer:2019lhp},
it is possible to imagine that the spacetime structure becomes smooth 
without loosing the (3+1)D expanding behavior
if one can approach $(s,k)=(0,0)$ in the large-$N$ limit.
Furthermore, the simulation supported 
a speculation that 
some classical solution
dominates at late times due to the expansion of space. 
This is important because it
shows the possibility to understand the late-time behavior of the model by
finding classical solutions that contain
a realistic cosmology \cite{Hanada:2005vr,Steinacker:2010rh,Chaney:2015ktw,Chaney:2015mfa,Steinacker:2016vgf,Chaney:2016npa,Stern:2018wud,Steinacker:2017vqw,Steinacker:2017bhb,Sperling:2019xar,Steinacker:2019dii}.
It is also expected that 
the solution that dominates at late times
can accommodate Standard Model particles as excitations around them. 
Early attempts to find such solutions 
used slightly modified models
by orbifolding \cite{Aoki:2002jt,Chatzistavrakidis:2010xi} or by toroidal 
compactification with a magnetic flux \cite{Aoki:2010gv,Aoki:2012ei}.
In Refs.~\cite{Chatzistavrakidis:2011gs,Nishimura:2013moa,
         Polychronakos:2013fma,Steinacker:2013eya,Steinacker:2014eua,Sperling:2018hys}, it was shown that the original
model can be used to realize intersecting D-branes and 
Refs.~\cite{Steinacker:2014fja,Aoki:2014cya,Honda:2019bdi} proposed matrix 
configurations that may correspond to phenomenologically 
viable low-energy effective theories.
For related work, see also 
Refs.~\cite{Chatzistavrakidis:2011su,Nishimura:2012rs}.

\subsection{the Euclidean version} \label{sec_EIKKT}
The Euclidean version of the IKKT matrix model, 
which we focus on in this paper, is obtained 
from the Lorentzian version by the Wick rotation
\begin{equation}
 A_{0} = i A_{10} \ , \quad \quad  \Gamma^{0} = - i \Gamma^{10} \  .
\label{wick_IKKT}
\end{equation}
The action 
is given by Eq.~\rf{IKKT_action}, 
where the contractions are now 
made with the metric $\delta_{\mu \nu}$ ($\mu,\nu=1,\ldots,10$),
and the partition function is defined by
\begin{equation}
 Z = \int dA d\psi\, e^{-S} = 
\int dA\, e^{-S_{\textrm{b}}}\, \textrm{Pf } {\cal M}\  ,
\label{EIKKT_partition}
\end{equation}
where the $16 (N^2-1) \times 16(N^2-1)$ anti-symmetric matrix ${\cal M}$ 
is defined by (\ref{M_definition}) with the replacement (\ref{wick_IKKT}). 
The Lorentz symmetry of the model
becomes an SO($10$) rotational symmetry acting on $A_\mu$ and $\psi_\alpha$.
Dynamical compactification of extra dimensions can be realized via 
the SSB of SO($10$) symmetry to SO($d$) with $d<10$.

The partition function \rf{EIKKT_partition} is 
finite \cite{Krauth:1998xh,Austing:2001pk}
despite the flat directions in the bosonic action $S_{\textrm{b}}$.
However, the Pfaffian  
$\textrm{Pf } {\cal M} = $ $ |\textrm{Pf } {\cal M}| e^{i \Gamma}$
is complex in general, which
causes a severe sign problem in numerical simulations.
The phase $\Gamma$ fluctuates wildly for large
matrices and plays a central role in the realization of the SSB of SO($10$).
In Ref.~\cite{Nishimura:2000ds,Nishimura:2000wf} 
it was shown that configurations 
with lower dimensions result in milder fluctuations of $\Gamma$, 
which points to the mechanism for favoring configurations 
less symmetric than SO($10$).

Monte Carlo simulations of the Euclidean IKKT and related matrix models 
have a long history.
Simplified versions can be defined by considering 
the reduction of the $D$-dimensional SYM theory to zero dimensions, 
which is possible for 
$D=3, 4, 6$ and $10$. 
The $D=3$ model is ill-defined because the partition
function is divergent \cite{Krauth:1998xh,Austing:2001pk}. The $D=4$ model
has a real non-negative fermion determinant in the effective action and 
Monte Carlo simulations
showed that the SO($4$) symmetry is not 
broken \cite{Ambjorn:2000bf,Ambjorn:2001xs}.
The $D=6$ model has a complex fermion determinant and simulations are
plagued by a severe sign problem as in the $D=10$ case.
Omitting the fermionic matrices \cite{Hotta:1998en} by simulating 
the bosonic model 
or omitting $\Gamma$ 
\cite{Ambjorn:2000dx,Anagnostopoulos:2013xga,Anagnostopoulos:2015gua} 
by simulating phase-quenched models, one finds no SSB.

Monte Carlo simulations including the complex phase were 
performed for the first time in 
Refs.~\cite{Anagnostopoulos:2001yb,Anagnostopoulos:2010ux,Anagnostopoulos:2011cn,Anagnostopoulos:2013xga,Anagnostopoulos:2015gua}. 
These calculations used a reweighting-based
method
\cite{Anagnostopoulos:2001yb,Anagnostopoulos:2010ux,Anagnostopoulos:2011cn}.
As an order parameter of the SSB, 
the eigenvalues $\lambda_\mu$
of the ``moment of inertia tensor''
\begin{equation}
 T_{\mu \nu} = \frac{1}{N} \textrm{tr} (A_{\mu} A_{\nu}) \label{t_munuE}
\end  {equation}
are defined with the ordering
\begin{equation}
\label{e01}
\lambda_1 > \lambda_2 > \ldots > \lambda_{10}\  .
\end{equation}
When the SSB of SO($D$) to SO($d$) with $d<D$ is realized, 
the expectation values
$\vev{\lambda_1}=$ $\vev{\lambda_2}=$ $\ldots=\vev{\lambda_d}$
are larger than 
the other $\vev{\lambda_\mu}$ 
($\mu=d+1,\ldots,D$).\footnote{Note that
the expectation values $\vev{\lambda_\mu}$ are calculated {\it after} 
the ordering \rf{e01}
and that equality is expected only in the large-$N$ limit.} 
The effect of the sign problem is reduced by simulating phase-quenched 
``microcanonical'' systems with
the constraints $\prod_\mu\delta(\lambda_\mu - x_\mu)$.
Here 
the values of $x_\mu$ are chosen appropriately
assuming that the SSB of SO($D$) to SO($d$) with $d<D$ is realized
for each $d$.
Varying the parameters $x_\mu$, one can sample 
in entropically highly suppressed regions of
the configuration space in which the fluctuations of $\Gamma$ are milder.
The expectation values $\vev{\lambda_\mu}$ for each SO($d$)
symmetric vacuum are obtained by minimizing the free energy for
the constrained system with respect to the parameters $x_\mu$
using the saddle-point approximation, which is justified at large $N$.
The effect of the phase is factorized
in the free energy,
and it can be obtained by
computing the average phase
as a function of $x_\mu$.
The results for $\vev{\lambda_\mu}$ obtained in this way
for each SO($d$) symmetric vacuum are found to be consistent with the results
obtained by the GEM to be discussed below.
However, comparison of the free energy for different vacua
requires integration over $x_\mu$,
which cannot be done
accurately enough to draw a definite conclusion on the SSB pattern
due to propagation of both systematic and statistical errors.




The issue of the SSB of SO($D$) was
also addressed analytically by using a
systematic expansion called the GEM.
Although the method involves only perturbative calculations,
it can give nonperturbative information
on the model to which it is applied \cite{Stevenson:1981vj},
as we review in more detail in Section \ref{sec_GEM}.
In the context of matrix models, the GEM has been first 
applied to the BFSS matrix theory \cite{Kabat:1999hp}
and to simplified versions of 
the Euclidean IKKT model \cite{Oda:2000im, Sugino:2001fn}.
Then the SSB of rotational symmetry 
in the Euclidean IKKT model and related models
has been investigated 
intensively \cite{Nishimura:2001sx,Kawai:2002jk,Nishimura:2002va,Kawai:2002ub,Nishimura:2003gz,Nishimura:2004ts,Aoyama:2006di,Aoyama:2006rk,Aoyama:2006je,Aoyama:2010ry,Nishimura:2011xy}.
One expands around a Gaussian action $S_0$
introduced by hand and containing many parameters, one for each quadratic term. 
In order to reduce the number of parameters, 
an ``ansatz'' that has an SO($d$) symmetry is considered.
The free energy and the
expectation values of observables are calculated in an expansion 
around $S_0$ 
as
functions of the parameters introduced. 
One can actually 
determine the region of the parameter space in which
the free energy is independent of the parameters,
and using these values of the parameters, one can obtain
the free energy 
and the observables $\vev{\lambda_\mu}$ for each $d$.

The $D=6$ model was studied in this way \cite{Aoyama:2010ry} and 
the SSB to SO($3$) was found. 
The free energy and the spacetime extent
were calculated
up to the fifth order 
for the SO($d$) ansatz with $2 \le d \le 5$,
and the free energy obtained for
the SO($3$) ansatz was found to be the minimum.
The extended directions
have an extent $R^2(d)=\vev{\lambda_\mu}$, $\mu=1,\ldots,d$, 
which is equal to $R^2(3)\approx 1.76$ for $d=3$.
The shrunken directions have an extent
$r^2=\vev{\lambda_\mu}\approx 0.223$
($\mu=d+1,\ldots,D$), which turned out to be almost independent of $d$.
Furthermore, the values $R^2(d)$ 
are such that they obey the constant volume property given by the relation
\begin{equation}
\label{e02}
R^d(d)\, r^{D-d} \approx l^D\, ,
\end{equation}
where $l$ is a length scale 
such that $v\equiv l^D$ gives the volume of spacetime.
Its value was calculated and found to be $l^2\approx 0.627$. These values
are consistent with the Monte Carlo simulations in
Refs.~\cite{Anagnostopoulos:2013xga,Anagnostopoulos:2017gos}.

A similar study was done also for the $D=10$
model \cite{Nishimura:2011xy}.
A systematic computation up to the third order 
was carried out for the SO($d$) ansatzes with $2\leq d \leq 7$. 
The free energy for the SO($3$) ansatz was found to be the minimum, 
suggesting also the SSB to SO($3$). 
The values of the large and small extents of space $R(d)$, $r$
were calculated and found to have similar properties 
as in the $D=6$ case.
In particular, 
one obtains
\begin{equation}
\label{e03}
R^2 (3) \approx 3.27 \ , \quad\quad
r^2 \approx 0.155 \ , \quad\quad
l^2 \approx 0.383 \ .
\end{equation}
These results are consistent with the ones obtained
by Monte Carlo simulations \cite{Anagnostopoulos:2015gua}.
Note also that these values are obtained in the large-$N$ limit 
and that they are finite.
This should be contrasted with the results 
in the Lorentzian model,
where space seems to expand indefinitely in the large-$N$ limit.

\section{Applying the CLM to the Euclidean IKKT model}
\label{sec:mc}


As we reviewed in the previous section,
the SSB of SO(10) symmetry in the Euclidean IKKT model
is expected to occur due to the effect of the phase of the Pfaffian
in (\ref{EIKKT_partition}), which causes the sign problem.
The aim of the present work is to 
use the CLM to overcome this sign problem,
and 
to understand the SSB pattern from first principles.



Here we 
apply the CLM to investigate the SSB of the SO($10$) symmetry 
in the Euclidean IKKT model
in a way similar
to Refs.~\cite{Ito:2016efb,Anagnostopoulos:2017gos}. 
We discuss how to probe the SSB
by using appropriate order parameters
and taking appropriate limits.
We also discuss important
techniques used to avoid known problems in the CLM
such as large excursions in the anti-Hermitian direction 
and the singular-drift 
problem \cite{Aarts:2009uq,Aarts:2011ax,Nishimura:2015pba,Nagata:2016vkn,Salcedo:2016kyy,Aarts:2017vrv}.
These techniques include a deformation of the fermionic action, the
adaptive stepsize and gauge cooling. 
By satisfying certain criteria \cite{Nagata:2016vkn,Nagata:2018net},
we can ensure that the
CLM yields correct 
results 
for a large space of parameters. 
The discussion is brief and more
details can be found in Ref.~\cite{Anagnostopoulos:2017gos}. 
Our main results are presented in the last subsection.

\subsection{the complex Langevin method}
\label{sec:cle}

The model \rf{EIKKT_partition}
we investigate can be written as
\begin{equation}
Z = 
\int dA\,{\rm e}^{-\Seff} \ ,
\label{m:1}
\end{equation}
where we define 
the effective action $\Seff=S_{\rm b}-\log\Pf{\cal M}$, which is complex.
In the CLM, we complexify the dynamical variables, which amounts
to regarding $A_\mu$ as general complex traceless matrices,
and consider their fictitious time evolution governed by
the complex Langevin equation, which is given as
\begin{equation}
\label{m:2}
\frac{d\left(A_\mu(t)\right)_{ij}}{dt} = 
-\frac{\partial \Seff[A_\mu(t)]}{\partial \left(A_\mu\right)_{ji}}
+\left(\eta_\mu\right)_{ij}(t) \  .
\end{equation}
Here $t$ is the fictitious time and $\eta_\mu(t)$ are 
traceless Hermitian matrices whose elements are random variables 
obeying the Gaussian distribution 
$\propto\exp\left(-\dfrac{1}{4}\int\tr\left\{\eta_\mu(t)\right\}^2\,dt\right)$. 
The first term on the right-hand side is called the drift term, which
is given as
\begin{equation}
\label{m:3}
 \frac{\partial\Seff    }{\partial \left(A_\mu\right)_{ji}}
=\frac{\partial S_{\rm b}     }{\partial \left(A_\mu\right)_{ji}}
-\frac{1}{2}\Tr\left(
{\cal M}^{-1} \frac{\partial {\cal M}}{\partial \left(A_\mu\right)_{ji}}
\right)\  ,
\end{equation}
where $\Tr$ represents the trace of a $16(N^2-1)\times 16(N^2-1)$
matrix. 
The expectation value of an observable ${\cal O}[A_\mu]$ 
can be calculated from
\begin{equation}
\label{m:4}
\vev{{\cal O}[A_\mu]} = \frac{1}{T}\int_{t_0}^{t_0+T}{\cal O}[A_\mu(t)]dt\  ,
\end{equation}
where $A_\mu(t)$ is a general complex matrix 
solution of \rf{m:2}, $t_0$ is the thermalization time 
and $T$ is large enough in order to obtain good statistics. 
Upon complexification of the matrices $A_\mu(t)$, 
the observable ${\cal O}[A_\mu(t)]$
depends on general complex matrices. 
The analyticity of the function
${\cal O}[A_\mu]$ plays a crucial role in the proof of the
validity of \rf{m:4}\cite{Aarts:2009uq,Aarts:2011ax,Nagata:2016vkn}.

The numerical solution of \rf{m:2} involves 
the discretization of the time $t$ given as
\begin{equation}
\label{m:5}
\left(A_\mu\right)_{ij}(t+\Delta t) = \left(A_\mu\right)_{ij}(t)
-\Delta t  \frac{\partial S[A_\mu(t)]}{\partial \left(A_\mu\right)_{ji}}
+\sqrt{\Delta t}\, \left(\eta_\mu\right)_{ij}(t)\  .
\end{equation}
The square root in the last term comes from the chosen
normalization of the $\eta_\mu(t)$ so that their probability
distribution is 
$\propto\exp\left(-\dfrac{1}{4}\sum_t\tr\left\{\eta_\mu(t)\right\}^2\right)$.

\subsection{how to probe the SSB}
\label{sec:opssb}

In order to probe the SSB, 
we break the SO($10$) symmetry explicitly by adding the terms
\begin{equation}
\label{m:7}
\Delta S_{\rm b} = \frac{1}{2} N \epsilonm 
 \sum_{\mu=1}^{10} m_\mu  \tr\left(A_\mu\right)^2
\end{equation}
to the action, where $0<m_1\leq\ldots\leq m_{10}$,
and take the $\epsilon \rightarrow 0 $ limit \emph{after} taking 
the large-$N$ limit.
As the order parameters, 
we consider \cite{Ito:2016efb,Anagnostopoulos:2017gos}
\begin{equation}
\label{m:6}
\lambda_\mu = \frac{1}{N} \tr (A_\mu)^2 \ , \quad
\mu = 1,\ldots,10\  ,
\end{equation}
where no sum over $\mu$ is 
taken\footnote{We avoid the use of the eigenvalues
of $T_{\mu\nu}$ in Eq.~\rf{t_munuE} so that we do not enter 
into the subtleties involved in the holomorphicity of the observables. 
We have measured them, however, using an ordering
based on their real part and the results are quantitatively 
identical to those obtained by using Eq.~\rf{m:6}.}.
The $\lambda_\mu$ are complex for
a generic complexified configuration $A_\mu(t)$, 
but after taking the average using
Eq.~\rf{m:4}, the $\vev{\lambda_\mu}$ become real. 
This is due to the
symmetry of the drift term \rf{m:3} under $A_\mu \longmapsto (A_\mu)^\dagger$
for $\mu=1,\ldots,9$ and $A_{10} \longmapsto -(A_{10})^\dagger$.
Due to the choice of the ordering of $m_\mu$, 
we have $\vev{\lambda_1}\geq \ldots \geq \vev{\lambda_{10}}$ for finite $N$. 
If there is no SSB of the SO($10$) rotational symmetry, 
all $\vev{\lambda_\mu}$ are equal in the $N\rightarrow \infty$
and $\epsilon \rightarrow 0$ limits. 
If it turns out that some of them are different, 
we conclude that SSB occurs.

\subsection{some techniques to make the CLM work}
\label{sec:gc}

In order for the CLM to yield correct results 
for the expectation value $\vev{{\cal O}[A_\mu]}$ of an observable
${\cal O}[A_\mu]$, 
the probability distribution $P(A^{({\rm R})}_\mu,A^{({\rm I})}_\mu;t)$ of the
general complex matrix solutions $A_\mu(t)$ to Eq.~\rf{m:2}, where
$A_\mu^{({\rm R})}(t)=(A_\mu(t)+A_\mu^\dagger(t))/2$, 
$A_\mu^{({\rm I})}(t)=(A_\mu(t)-A_\mu^\dagger(t))/2i$, 
must satisfy the relation 
\begin{equation}
\label{m:8}
\int dA_\mu\, \rho(A_\mu;t) {\cal O}[A_\mu] =
\int dA_\mu^{({\rm R})}dA_\mu^{({\rm I})}\, P(A_\mu^{({\rm R})},A_\mu^{({\rm I})};t) 
     {\cal O}[A_\mu^{({\rm R})}+i A_\mu^{({\rm I})}]\  .
\end{equation}
On the left-hand side of the above equation, $A_\mu$ are the original Hermitian
matrices in the model \rf{m:1} and $\rho(A_\mu;t)$ is a complex
weight which is a solution to a Fokker-Planck equation such that
$\lim\limits_{t\to\infty}\rho(A_\mu;t) = {\rm e}^{-\Seff[A_\mu]}/Z$, 
giving the desired
$\vev{{\cal O}[A_\mu]}$ in the $t\to\infty$ limit 
(for details, see e.g., Ref.~\cite{Nagata:2016vkn}).
The right-hand side involves the (real positive)
probability distribution of the complex matrix solutions of the
complex Langevin equation \rf{m:2} and the analytic continuation of
${\cal O}[A_\mu] \longmapsto {\cal O}[A_\mu^{({\rm R})}+i A_\mu^{({\rm I})}]$,
and in the $t\to\infty$ limit, it essentially gives the quantity
on the right-hand side of Eq.~\rf{m:4}. 
A sufficient condition for Eq.\rf{m:8}
to hold is that the probability distribution $p(u)$ of
the magnitude of the drift\footnote{The effective action $\Seff[A_\mu]$ 
in Eq.~(\ref{m:9}) should be modified appropriately
taking account of the terms (\ref{m:7}) and (\ref{m:11}) 
added to the action in actual simulation.}
\begin{equation}
\label{m:9}
u = \sqrt{\frac{1}{10 N^3}\sum_{\mu=1}^{10} \sum_{i,j=1}^N
\left| \frac{\partial \Seff\left[A_\mu\right]}{\partial \left(A_\mu\right)_{ij}}  \right|^2 } 
\end{equation}
in the ensemble defined by
$P(A^{({\rm R})},A^{({\rm I})};t)$ 
falls off exponentially or faster\cite{Nagata:2016vkn}.
The above condition can be violated if the $A_\mu(t)$ makes long excursions
in the anti-Hermitian direction (``excursion problem''). In order to
avoid it, we employ gauge cooling \cite{Seiler:2012wz} 
in our simulations by minimizing 
the ``Hermiticity norm'' defined by 
\begin{equation}
\label{m:10}
{\cal N}_{\rm H} = - \frac{1}{10 N} \sum_{\mu=1}^{10} 
 \tr \left\{ \left(A_\mu - A_\mu^\dagger \right)^2 \right\} 
\end{equation}
at each step of (\ref{m:5}).
(See Ref.~\cite{Anagnostopoulos:2017gos} for more details.)
It was proven \cite{Nagata:2015uga,Nagata:2016vkn} 
that adding the gauge cooling procedure
in the CLM does not affect the argument for its justification.

The stability of the evolution given by Eq.~\rf{m:5} is also controlled
by the adaptive stepsize \cite{Aarts:2009dg}. 
The typical magnitude of the drift term
$|\partial S /\partial (A_\mu)_{ji}| \Delta t \sim u \sqrt{N} \Delta t$ 
in \rf{m:5}
has to be kept small compared with the typical 
$|(A_\mu)_{ij}| \gtrsim (\lambda_{\mbox{\scriptsize min}}/N)^{1/2}$,
where
$\lambda_{\mbox{\scriptsize min}}\equiv\min\limits_\mu |\lambda_\mu|$ for a given 
configuration. Therefore, at each step we adjust 
$\Delta t = $
$\min\{ \Delta t_{\mbox{\scriptsize max}},$
$\epsilon_t\left(\sqrt{\lambda_{\mbox{\scriptsize min}}}/(N u)\right)\}$,
where $\Delta t_{\mbox{\scriptsize max}}$ and $\epsilon_t$
are tuned so that the results have negligible finite $\Delta t$ errors.
Measurements are taken at fixed intervals of the Langevin time. 

Another reason for 
the condition on the drift distribution to be violated
is the singular-drift problem. 
This problem occurs because of the appearance of 
${\cal M}^{-1}$ in Eq.~\rf{m:3}
when the eigenvalues of $\cal M$ accumulate densely near zero. 
In order to avoid this problem, we
deform the fermionic action by adding a term
\begin{equation}
\label{m:11}
\Delta S_{\rm f} = -i \mf \frac{N}{2} \, \tr\left(
\psi_\alpha 
({\cal C}\Gamma_8 \Gamma_9^\dagger \Gamma_{10})_{\alpha\beta}
\psi_\beta
\right)\  ,
\end{equation}
where $\mf$ is the deformation parameter. 
The addition of this term
shifts the eigenvalue distribution of $\cal M$ away from the origin,
enabling us to avoid the singular-drift problem. 
This technique was proposed originally in Ref.~\cite{Ito:2016efb},
where the singular-drift problem was indeed 
overcome in an SO($4$) symmetric matrix model
with a complex fermion determinant, and it was used 
successfully also in the $D=6$ version
of the IKKT matrix model \cite{Anagnostopoulos:2017gos}. 
Note that the above term (\ref{m:11}) breaks
the SO($10$) symmetry down to ${\rm SO}(7)\times {\rm SO}(3)$ explicitly. 
Therefore, we
investigate whether the remaining SO($7$) symmetry breaks down to smaller
subgroups as $\mf$ is varied and discuss what occurs
at $\mf = 0$.

Note that as $\mf\to\infty$, the fermionic degrees of freedom
decouple and we obtain the dimensionally reduced
Yang-Mills model without fermionic matrices (``bosonic model''), 
which is SO($10$) symmetric \cite{Anagnostopoulos:2015gua}. 
Thus the deformation \rf{m:11} can be regarded as an
interpolation between the IKKT model and the bosonic model. 

The presence of the singular-drift problem depends 
on the values of
the parameters $\mf$ 
and $\epsilonm$.
For large enough $\mf$ the problem is cured, but it reappears as $\mf$ is
reduced for values of $\epsilonm$ smaller than some value depending on
$\mf$.
In our simulations, we choose the range of the
simulated $(\mf,\epsilonm)$ for each $N$ carefully 
so that the probability distribution of
the magnitude of the drift \rf{m:9} falls off faster than
exponentially, ensuring that we do not have the singular-drift problem.
In Appendix \ref{sec:large-N-singular-drift},
we show some results suggesting that
the singular-drift problem vanishes for large enough $N$
for given values of $(\mf,\epsilonm)$.

\subsection{results of the CLM}
\label{sec:results}

In this subsection we present our main results obtained 
in the way discussed in the previous subsections. 
The deformation \rf{m:11} breaks
the SO($10$) symmetry down to 
${\rm SO}(7)\times {\rm SO}(3)$,
and we examine if the SO($7$) symmetry is broken down to a smaller group 
for $\mf = 3.0$, 1.4, 1.0, 0.9, 0.7
to consider what happens for the undeformed IKKT model \rf{m:1}
corresponding to $\mf = 0$.
In order to probe the SSB, one has to take the $N\to\infty$ limit 
first and then the $\epsilonm\to 0$ limit. 
  
The $m_\mu$ in Eq.~\rf{m:7} are chosen so that this term 
does not break SO($10$) completely because
otherwise the spectrum of $m_\mu$ becomes too wide
to make the $\epsilonm\to 0$ extrapolation reliably.
For $\mf=3.0$, 
we choose $m_\mu=(0.5, 0.5, 0.5, 1, 2, 4, 8, 8, 8, 8)$,
which enables us to distinguish SO($d$) vacua with $d=3,4,5,6,7$.
For smaller values of $\mf$, 
we choose $m_\mu=(0.5, 0.5, 1, 2, 4, 8, 8, 8, 8,8)$,
which enables us to distinguish SO($d$) vacua with $d=2,3,4,7$.
In particular, we may confirm that the SO(3) symmetry remains unbroken
by seeing that $\langle \lambda_1  \rangle
= \langle \lambda_2 \rangle$ and $ \langle \lambda_3 \rangle$ agree
in the $N \rightarrow \infty$ and $\epsilonm \rightarrow 0$ limits.
On the other hand, this choice of $m_\mu$  has a drawback that
$\lambda_6$ and $\lambda_7$ are mixed up because of $m_6 = m_7$,
and hence one cannot distinguish SO(5) and SO(6) vacua.
This does not cause any harm, however, as far as we find that
$\langle \lambda_4 \rangle$ and 
$\langle \lambda_5 \rangle$ do not agree
in the $N \rightarrow \infty$ and $\epsilonm \rightarrow 0$ limits,
which implies that the SO(7) symmetry is broken to 
SO(4) or lower symmetries.



\begin{figure}[t]
\centering 
\includegraphics[width=0.7\textwidth]{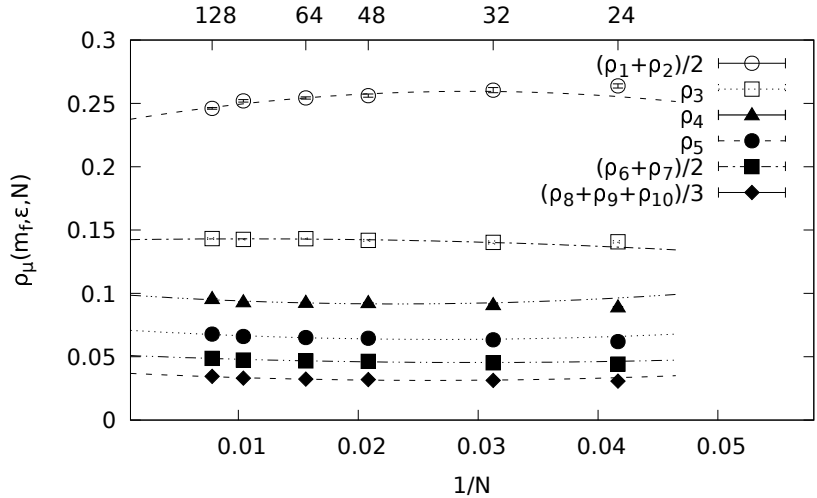}
    \caption{The large-$N$ extrapolation of $\rho_\mu(\mf, \epsilonm,N)$  
for $\mf=1.0$, $\epsilonm=0.2$ 
with $m_\mu=(0.5, 0.5, 1, 2, 4, 8, 8, 8, 8, 8)$.
The $\rho_\mu(\mf, \epsilonm,N)$ are averaged 
for $\mu=1,2$, $\mu=6,7$ and $\mu=8,9,10$ 
to increase statistics.
A quadratic fit with respect to $1/N$ is performed.
\label{f:r01}}
\end{figure}

For given $m_\mu$ in Eq.~\rf{m:7}, the large-$N$ limit is obtained by
first 
computing the ratio
\begin{equation}
\label{r01}
 \rho_\mu(\mf, \epsilonm,N) = 
  \frac{\vev{\lambda_\mu}_{\mf, \epsilonm,N}}
{\sum_{\nu=1}^{10}\vev{\lambda_\nu}_{\mf, \epsilonm,N}}\  ,
\end{equation}
and then by making a large-$N$ extrapolation
\begin{equation}
\label{r02}
 \rho_\mu(\mf, \epsilonm)= \lim_{N\to \infty} \rho_\mu(\mf, \epsilonm,N)\ .
\end{equation}
The reason for investigating the ratio (\ref{r01})
instead of $\vev{\lambda_\mu}$ is 
that a large part of the $\epsilonm$ dependence is canceled 
between the numerator and the denominator,
which makes the $\epsilonm\to 0$ extrapolation more reliable. 
The large-$N$ extrapolation is performed 
by plotting $\rho_\mu(\mf,\epsilonm,N)$ 
against $1/N$ and making 
a quadratic fit with respect to $1/N$.
We find that the quadratic 
term
is necessary in the fits
especially for small $(\mf,\epsilonm)$.
In Fig.~\ref{f:r01} we show a typical case of such a fit. 

\begin{figure}[htbp]
\centering 
\includegraphics[width=0.45\textwidth]{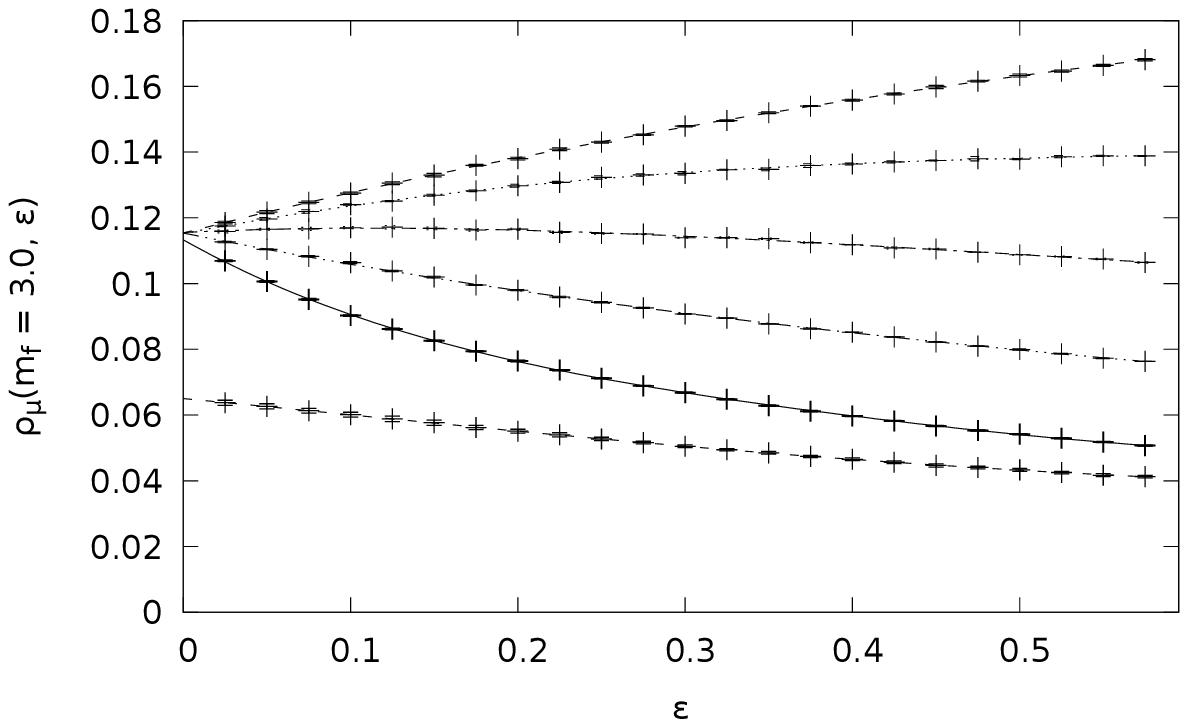} 
\includegraphics[width=0.45\textwidth]{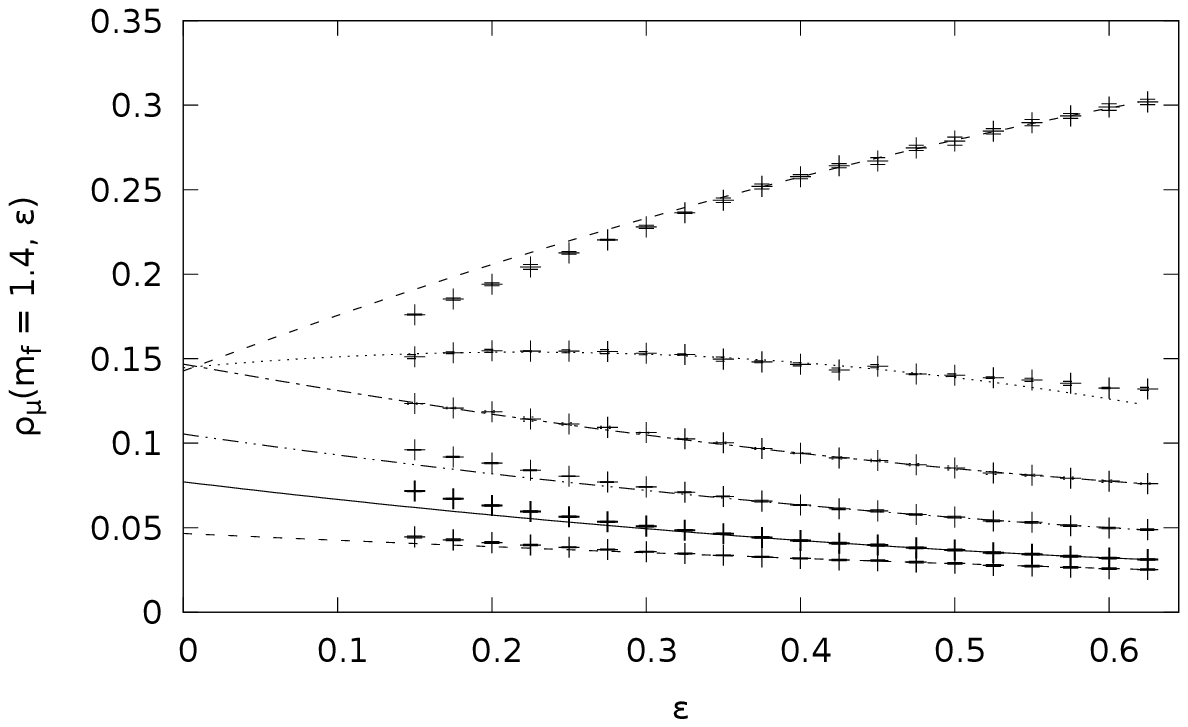}
\includegraphics[width=0.45\textwidth]{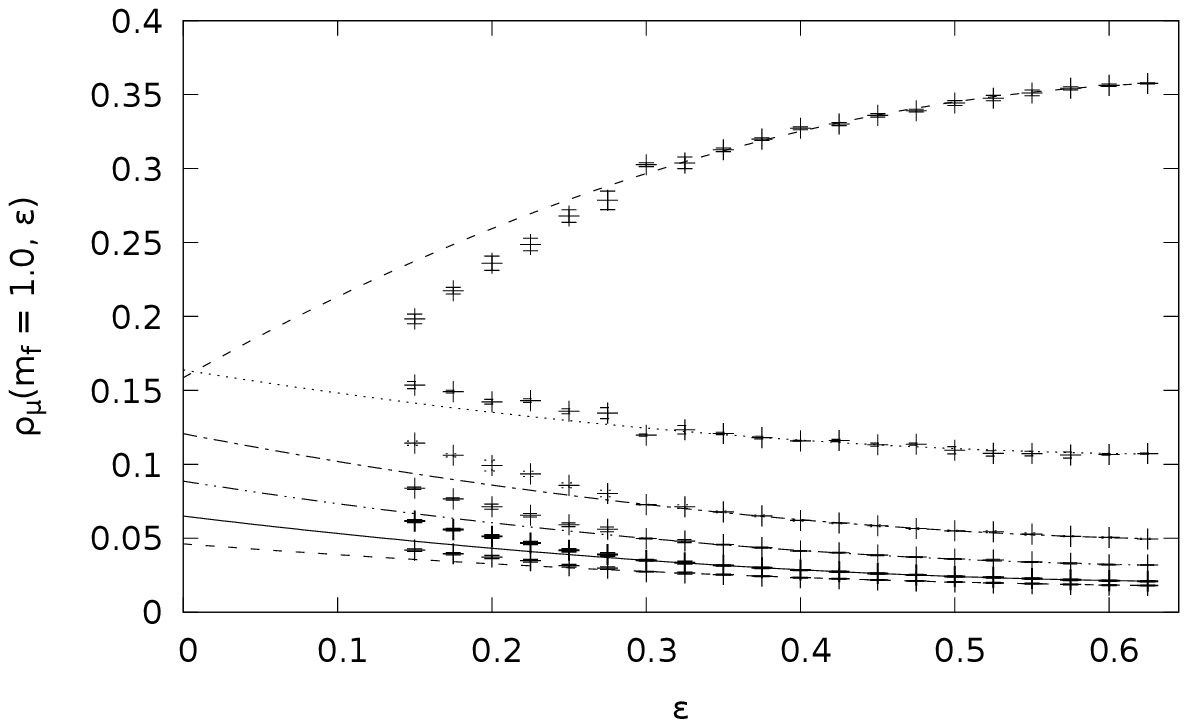}
\includegraphics[width=0.45\textwidth]{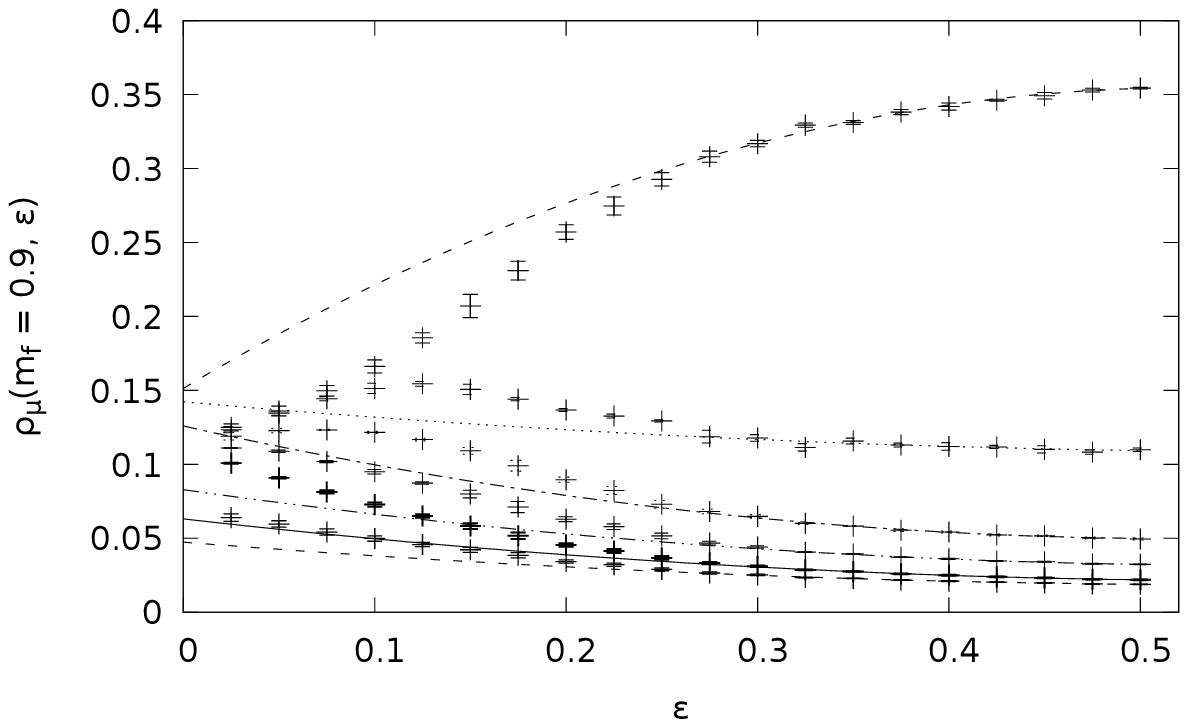}
\includegraphics[width=0.45\textwidth]{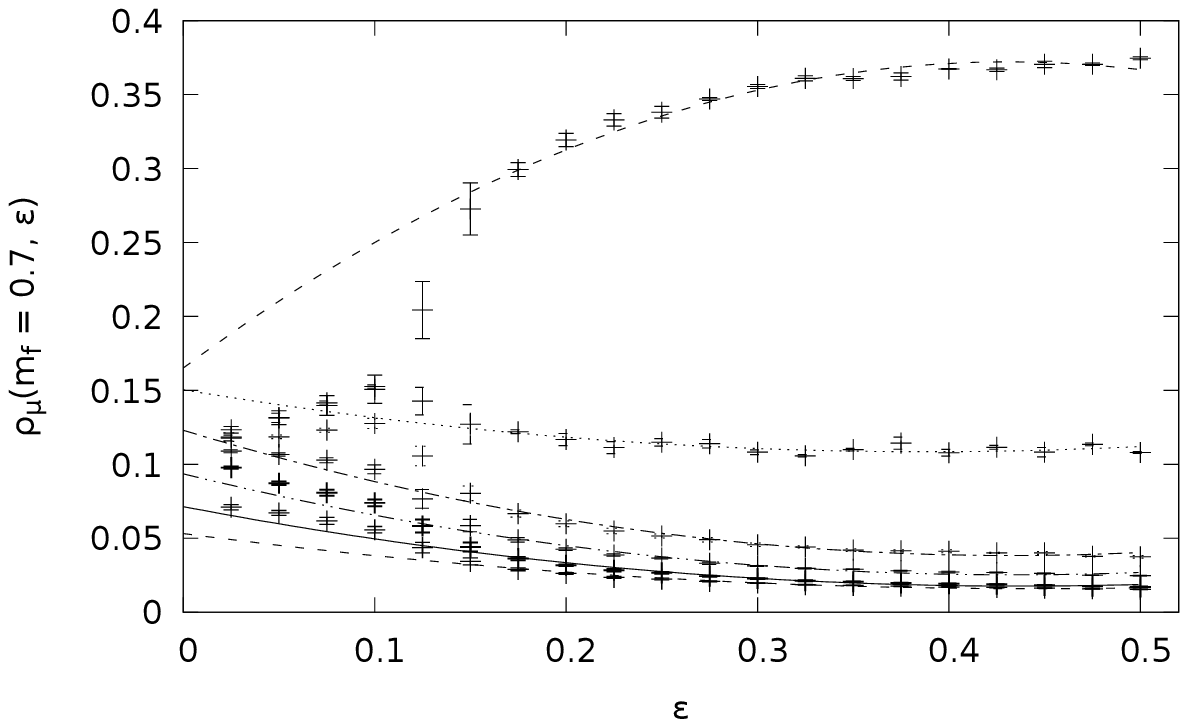}
    \caption{The $\rho_\mu(\mf,\epsilonm)$ in Eq.~\protect\rf{r02} 
are plotted against $\epsilonm$ for $\mf=3.0$ (Top-Left), 
$\mf=1.4$ (Top-Right), $\mf=1.0$ (Middle-Left),
$\mf=0.9$ (Middle-Right) and $\mf=0.7$ (Bottom). 
We use
$m_\mu=(0.5, 0.5, 0.5, 1, 2, 4, 8, 8, 8, 8)$
for $\mf=3.0$ and $m_\mu=(0.5, 0.5, 1, 2, 4, 8, 8, 8, 8, 8)$
for the other values of $\mf$.
The continuous lines are polynomial fits in $\epsilonm$. 
For $\mf=3.0$ a quartic fit is performed, 
whereas for the other values of $\mf$ the
fits are quadratic in $\epsilonm$. 
In the $\mf=3.0$ plot, the curves from top to bottom are 
$(\rho_1+\rho_2+\rho_3)/3$, $\rho_4$, $\rho_5$, 
$\rho_6$, $\rho_7$  and $(\rho_8+\rho_9+\rho_{10})/3$. 
For the other plots, the curves from top to bottom are 
$(\rho_1+\rho_2)/2$,  $\rho_3$, $\rho_4$, $\rho_5$, $(\rho_6+\rho_7)/2$  
and $(\rho_8+\rho_9+\rho_{10})/3$.
\label{f:r02}}
\end{figure}

In Fig.~\ref{f:r02}
we plot the large-$N$ extrapolated values 
$\rho_\mu(\mf,\epsilonm)$ as a function of $\epsilonm$ for
$\mf=3.0$, 1.4, 1.0, 0.9 and $0.7$.
In order to increase statistics, 
we average the  $\rho_\mu(\mf,\epsilonm)$ for the 
$\mu$ that we expect to give equal values due to 
degeneracies in $m_\mu$. 
For $\mf=3.0$, we use $m_\mu=(0.5,0.5,0.5,1,2,4,8,8,8,8)$,
which implies that 
the $\rho_\mu(\mf,\epsilonm)$ should be equal 
for $\mu=1,2,3$ and  $\mu=8,9,10$.
For $\mf=1.4, 1.0, 0.9, 0.7$,
we use $m_\mu=(0.5,0.5,1,2,4,8,8,8,8,8)$, 
which implies that
the $\rho_\mu(\mf,\epsilonm)$ should be equal for
$\mu=1,2$, $\mu=6,7$ and $\mu=8,9,10$. 
The $\epsilonm\to 0$ extrapolation
\begin{equation}
\label{r03}
\rho_\mu(\mf) = \lim_{\epsilonm\to 0} \rho_\mu(\mf,\epsilonm)
\end{equation}
is performed by fitting $\rho_\mu(\mf,\epsilonm)$ to 
a polynomial in $\epsilonm$.


From the extrapolated values $\rho_\mu(\mf)$,
we find that the SO($7$) symmetry
of the deformed model is not spontaneously broken
at $\mf = 3.0$, but it is actually broken to SO($4$)
for $\mf = 1.4$ and to SO($3$) for $\mf = 1.0$, $0.9$, $0.7$.
Thus as $\mf$ is decreased,
the SO($7$) symmetry seems to be spontaneously broken 
to smaller subgroups gradually in the
same way as it was observed in the $D=6$ 
case \cite{Anagnostopoulos:2017gos}. 
However, we consider that the symmetry is not going to be broken
further down to SO($2$) at smaller $\mf$.
This is based on the fact that
the Pfaffian vanishes identically for
strictly 2D configurations \cite{Nishimura:2000ds,Nishimura:2000wf},
which implies that the mechanism of SSB due to the phase
of the Pfaffian no longer works there\footnote{This is also reflected in 
the GEM results \cite{Aoyama:2010ry,Nishimura:2011xy}
for the free energy of the SO($d$) vacuum,
which becomes much larger for $d=2$ than for $d\ge 3$.}.
Hence our results are 
consistent with the results obtained by the GEM
for the undeformed model,
which show that the SO(3) vacuum has the smallest free energy.

When we make the fit for the $\epsilonm\to 0$ extrapolation
in Fig.~\ref{f:r02},
we excluded the data points at small $\epsilonm$ except for $\mf =3.0$.
These data points show a clear tendency towards the restoration
of SO(10) rotational symmetry.
We consider that this is due to the insufficient
large-$N$ extrapolation performed
to obtain these points due to severe finite-$N$ effects for 
small $\mf$ and small $\epsilonm$.
As we discussed earlier, the parameter $\mf$ 
interpolates between the bosonic and supersymmetric models. 
In the bosonic model
there is a strong attractive force between all the pairs
of the $N$ spacetime points 
defined by the eigenvalues of the $A_\mu$, which represents
an O($N^2$) effect \cite{Hotta:1998en}, 
whereas in the supersymmetric model the attractive force is mostly
canceled due to supersymmetry
and it becomes an O($N$) effect \cite{Aoki:1998vn}. 
Also, a large value of $\epsilonm$ reduces the spacetime volume
according to Eq.~\rf{m:7}. 
Therefore, for given $N$ the density of spacetime points 
becomes small as one takes smaller values of $\mf$ and $\epsilonm$,
and this is the reason why finite-$N$ effects become severe in this region.

When finite-$N$ effects are important, SSB is suppressed and configurations
appear symmetric.
The transition between the regions 
in which one sees true SSB effects and a (falsely) symmetric behavior 
can be seen as a clear smooth crossover in the three lower plots 
in Fig.~\ref{f:r02} for $\mf=0.7, 0.9$ and $1.0$. 
The crossover disappears for larger values of $\mf$ as expected. 
This is seen to happen already for $\mf=3.0$ in Fig.~\ref{f:r02}, 
whereas the crossover is very mild for $\mf=1.4$.
The fitting region is chosen so that 
finite-$N$ effects in the small-$\epsilonm$ region
do not affect the $\epsilonm\to 0$ extrapolation.

\section{Consistency check based on the GEM} \label{sec_GEM}

In the previous section we discussed the pattern of the SSB of SO($10$)
as the deformation parameter $\mf$ in Eq.~\rf{m:11} is varied. As it has
already been mentioned, $\mf$ interpolates between the IKKT
matrix model ($\mf\to 0$), where SSB to SO($3$) is expected to occur,
and the bosonic model ($\mf\to\infty$), where there is no SSB. 
As the value of $\mf$ is decreased, the SSB pattern gradually 
changes from more symmetric vacua to less symmetric ones.
However, we had to make extrapolations,
first the large-$N$ extrapolation and then
the $\epsilonm\to 0$ extrapolation,
which introduce systematic errors.
Therefore, it is important to verify the
results using a completely different approach.

In Section \ref{sec_EIKKT} we made a short introduction to the GEM
and its application to the Euclidean IKKT matrix model. 
In particular, SO($10$) 
was found to be broken down to SO($3$) \cite{Nishimura:2011xy}.
Here we apply the GEM to the
IKKT matrix model deformed by 
the parameter $\mf$.
We perform a three-loop calculation using SO($d$)
symmetric ansatzes, where $d=6, 7$,
and calculate the free energy.
We observe a trend that the free energy for the SO(6) vacuum
becomes smaller than the SO(7) vacuum as we decrease $\mf$.
We also find that the extent of space obtained at
$\mf = 3.0$ agrees very well between the two methods.



It should be noted that the GEM does not depend on the large-$N$ and
small-$\epsilonm$ extrapolations. 
The $N\to\infty$ limit can be taken by simply using planar graphs, 
and the SSB can be readily probed by comparing the free energy obtained
with various ansatzes for the Gaussian action with different symmetries.
Furthermore, the systematic errors of the GEM 
come mainly from
the truncation of the expansion and the determination of the parameters
giving the physical solutions,
which are completely different from the systematic errors 
of Monte Carlo simulations. 
Therefore the results of the two methods can be considered to
be independent and their consistency provides more confidence 
in our conclusion.

\subsection{applying the GEM to the deformed model}

Here we review the GEM used in Ref.~\cite{Nishimura:2011xy}
to study the original IKKT matrix model
and apply it to the deformed model
with the fermionic mass-like term (\ref{m:11}).
The basic idea is to introduce a Gaussian action $S_{0}$
and to rewrite the original action $S=\Sb+\Sf+\Delta \Sf$ as
\begin{equation}
S= S_{0}+ (S-S_{0})\ ,
\label{eq:01}
\end{equation}
where the first and second terms are regarded as the ``classical
action'' and the ``one-loop counterterm,'' respectively. 
This enables us to perform a perturbative expansion. 
The Gaussian action contains a large number of arbitrary parameters 
and the results depend on these parameters in general. 
On the other hand, physical
quantities should not depend on these parameters.
It turns out to be possible to find a range of these parameters 
in which observables are (almost) independent of their values. 

Let us consider the Gaussian action
\begin{equation}
\label{g:01}
S_{0} = \frac{N}{2}\sum_{\mu=1}^{10}M_{\mu}\mathrm{tr}(A_{\mu}^{2})
      + \frac{N}{2}\sum_{\alpha,\beta=1}^{16}\mathcal{A}_{\alpha\beta}
           \mathrm{tr}(\psi_{\alpha}\psi_{\beta}) \ ,
\end{equation}
where $M_{\mu}$ and $\mathcal{A}_{\alpha\beta}$
are the parameters for the bosonic part and
the fermionic part of the Gaussian action, respectively. 
We choose the ordering $0<M_1\leq\ldots\leq M_{10}$ so that
the notation matches with (\ref{m:7}).
The complex $16\times16$ antisymmetric matrix $\mathcal{A}_{\alpha\beta}$ 
in (\ref{g:01})
can be expanded using the 10-dimensional gamma matrices $\Gamma_{\mu}$ as
\begin{equation}
\label{g:02}
\mathcal{A}_{\alpha\beta}  = 
\sum_{\mu,\nu,\rho=1}^{10}\frac{i}{3!}m_{\mu\nu\rho}
({\cal C}\Gamma_{\mu}\Gamma_{\nu}^{\dagger}\Gamma_{\rho})_{\alpha\beta} \  ,
\end{equation}
where $m_{\mu\nu\rho}$ is a totally antisymmetric 3-form.
The effect of the fermion mass term (\ref{m:11}) can
be readily implemented by 
replacing the coefficient 
as $m_{8,9,10} \mapsto m_{8,9,10} - \mf$.

The partition function can
be rewritten as
\begin{eqnarray}
\label{g:03}
Z     & = & Z_{0}\langle e^{-(S-S_{0})}\rangle_{0} \  , \nonumber
\\
Z_{0} & = & \int dAd\psi\,e^{-S_{0}} \  , \nonumber
\end{eqnarray}
where $\langle\,\cdot\,\rangle_{0}$ represents the expectation value with
respect to the partition function for the Gaussian action $Z_{0}$.
One can expand the free energy $F=-\log Z$ perturbatively as
\begin{eqnarray}
F     & = & \sum_{k=0}^{\infty}f_{k} \ ,\label{eq:fe}\\
f_{0} & = & -\log Z_{0} \  , \nonumber\\
f_{k} & = & -\sum_{l=0}^{k}\frac{(-1)^{k-l}}{(k+1)!}{}_{k+l}
\mathrm{C}_{k-l}\langle(S_{\mathrm{b}}-S_{0})^{k-l}
(S_{\mathrm{f}})^{2l}\rangle_{\mathrm{C},0} \  , \nonumber
\end{eqnarray}
where the subscript C in $\langle\,\cdot\,\rangle_{\mathrm{C},0}$
means that only connected diagrams are summed over. 
Similarly, the expectation values of observables are given by
\begin{eqnarray}
\langle\mathcal{O}\rangle & = & 
\langle\mathcal{O}\rangle_{0}+\sum_{k=1}^{\infty}\mathcal{O}_{k}\  ,
\label{obs-expand}  \\
\mathcal{O}_{k} & = & 
\sum_{l=0}^{k}\frac{(-1)^{k-l}}{(k+l)!}{}_{k+l}\mathrm{C}_{k-l}
\langle\mathcal{O}(S_{\mathrm{b}}-S_{0})^{k-l}
(S_{\mathrm{f}})^{2l}\rangle_{\mathrm{C},0} \ . 
\nonumber
\end{eqnarray}
In practice, we truncate the infinite series 
such as (\ref{eq:fe}) and (\ref{obs-expand})
at some finite order and evaluate each term using Feynman diagrams,
where we restrict ourselves to planar diagrams since
we are interested in the large-$N$ limit.

As we already mentioned above, for a generic set of parameters
$M_{\mu}$ and $\mathcal{A_{\alpha\beta}}$, the results obtained in this way
depend on their values.
In order to find ``physical'' results that do not 
depend on these parameters,
we search for stationary points of the free energy
with respect to those parameters by solving the 
``self-consistency equations'' 
\begin{equation}
\frac{\partial}{\partial M_{\mu}}F=0 \ , \quad
\frac{\partial}{\partial m_{\mu\nu\rho}}F=0 \ .
\label{eq:self-con}
\end{equation}
The solutions to these equations 
are obtained numerically, and
they are used as the probe of the plateaus
for the values of the free energy in the space of the 
parameters $M_{\mu}$ and $\mathcal{A_{\alpha\beta}}$.


In order to consider the SO($d$) symmetric vacuum,
we impose the SO($d$) symmetry on the Gaussian action 
by setting $M_{1}=\ldots=M_{d}$
and $m_{\mu\nu\rho}=0$ unless $\mu,\nu$ and $\rho$ are different from
each other with $\mu,\nu,\rho\geq d+1$. 
Thus we are left with $(11-d)$ parameters from $M_{\mu}$ 
and $_{10-d}\mathrm{C}_{3}$ parameters from $m_{\mu\nu\rho}$.
However, it turns out hard to solve the self-consistency
equation (\ref{eq:self-con}) with more than 5 parameters. 
In the previous work,
we reduced the number of parameters by imposing some 
discrete symmetries $\Sigma_{d}$ to the shrunken directions as well,
where ${\rm SO}(d) \times \Sigma_{d}$ are subgroups of SO($10$).
In the present work, 
we focus on two cases, in which
we impose SO($d$)$\times Z_{3}$
with $d=6$ and $d=7$,
where $Z_{3}$ represents
a group of cyclic permutations of the 8th, 9th and 10th directions.
Note that the imposed symmetries are subgroups of
SO(7)$\times$SO(3), which remains after adding the mass term (\ref{m:11}).

\subsection{results of the GEM}


\begin{figure}
\centering{}\includegraphics[width=0.6\textwidth]{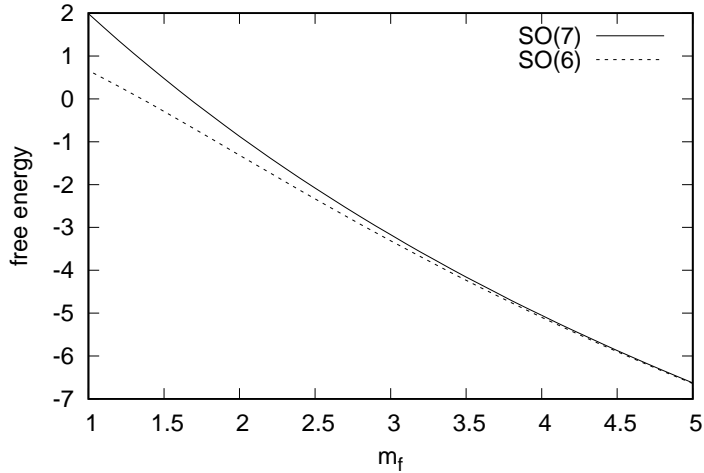}
\caption{The free energy calculated up to three loops for the 
solutions found with the 
SO($7$) and SO($6$) 
ansatzes
are plotted against the fermion mass $\mf$. 
We observe a clear tendency that the SO(6)
symmetric vacuum is more favored
as $\mf$ is decreased, whereas the free energy for the two ansatzes
tends to be degenerate as $\mf$ is increased.
\label{fig:free_energy_gem}}
\end{figure}

In Fig.~\ref{fig:free_energy_gem}
we plot the free energy calculated up to three loops for the 
solutions found with the 
SO($7$) and SO($6$) 
ansatzes
against the fermion mass $\mf$. 
We observe a clear tendency that the SO(6) 
symmetric vacuum is more favored
as $\mf$ is decreased.
However, the free energy for the two ansatzes
tends to become degenerate as $\mf$ is increased.
In this situation
it is difficult to identify the critical point,
given the accuracy of the GEM results.

\begin{figure}
\centering 
 \includegraphics[width=0.45\textwidth]{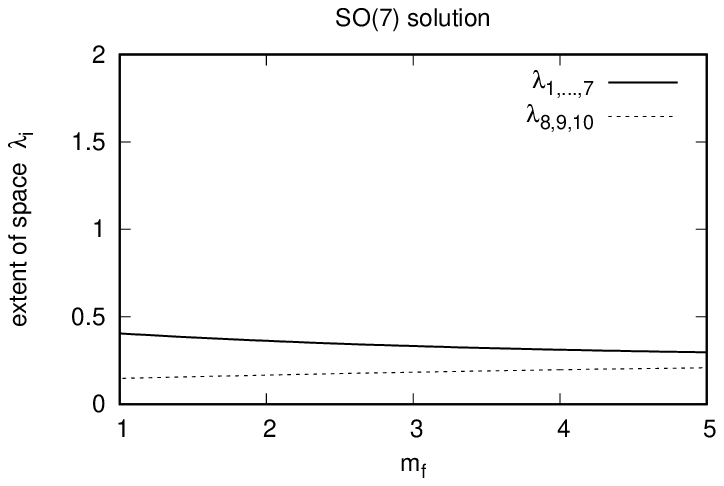}
 \includegraphics[width=0.45\textwidth]{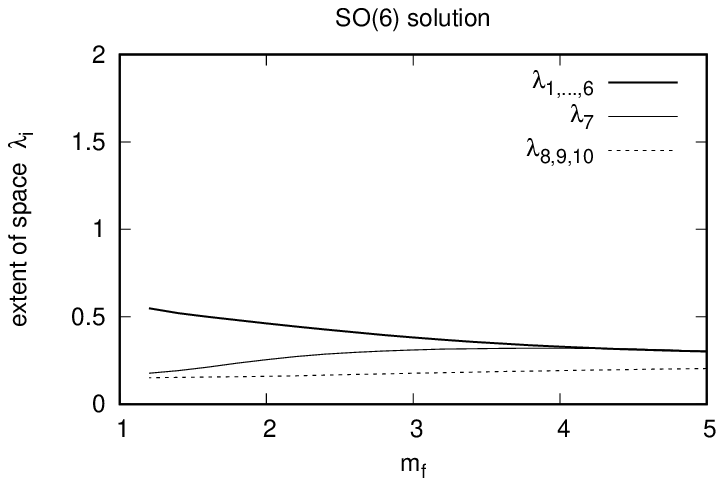}
\caption{The extent of space 
in each direction is plotted 
against $\mf$ 
for the SO(7) ansatz (Left) and 
for the SO(6) ansatz (Right).
For the SO(7) ansatz, we plot 
$\lambda_1  = \ldots = \lambda_7$
and $\lambda_8 = \lambda_9 = \lambda_{10}$.
For the SO(6) ansatz, we plot 
$\lambda_1  = \ldots = \lambda_6$,
$\lambda_7$
and $\lambda_8 = \lambda_9 = \lambda_{10}$.
\label{fig:extent_so4_so6}
}
\end{figure}

In Fig.~\ref{fig:extent_so4_so6},
we plot the extent of space 
$\lambda_i$ ($i=1,\ldots,10$) in each direction against $\mf$
for the SO(7) and SO(6) ansatzes.
For the SO(7) ansatz (Left), we plot $\lambda_1  = \ldots = \lambda_7$,
and $\lambda_8 = \lambda_9 = \lambda_{10}$.
We find that the two lines come close to each other,
which is consistent with the fact that the deformed model
becomes the bosonic model at $\mf = \infty$, 
where the full SO(10) symmetry is expected to be restored.
For the SO(6) ansatz (Right),
we plot $\lambda_1  = \ldots = \lambda_6$,
$\lambda_7$ and $\lambda_8 = \lambda_9 = \lambda_{10}$.
We find that the line in the middle corresponding to
$\lambda_7$ goes up as $\mf$ increases
and asymptotes to the line at the top corresponding to
$\lambda_1  = \ldots = \lambda_6$.
As a result, the plot looks almost identical to the plot for
the SO(7) ansatz in the large-$\mf$ region.
This clearly explains why the free energy for the two ansatz
asymptotes to each other as $\mf$ increases.
On the other hand, the line in the middle
goes down as $\mf$ decreases
and it seems to asymptote to the line at the bottom corresponding to
$\lambda_8  = \lambda_9 = \lambda_{10}$.
This is consistent with the GEM results for the undeformed model,
which suggest that the shrunken directions
have a common extent. It is also natural from the viewpoint that
the explicit breaking of SO(10) symmetry to ${\rm SO}(7) \times {\rm SO}(3)$
by the fermion mass term is removed as $\mf$ decreases.

The results of the CLM
at $\mf = 3.0$ shows that we obtain an SO(7) symmetric vacuum
with the extent of space given by
\begin{eqnarray}
\rho_1 = \cdots = \rho_7 = 0.115  \ , \quad
(\rho_8 + \rho_9 + \rho_{10})/3 = 0.065  \ .
\label{extent_CLM}
\end{eqnarray}
On the other hand, 
from the results plotted in Fig.~\ref{fig:extent_so4_so6} (Left),
we find for the SO(7) ansatz that
the extent of space is
$\lambda_1 = \ldots = \lambda_7 = 0.333$,
$\lambda_8 = \lambda_9 = \lambda_{10} = 0.184$ at $\mf = 3.0$.
Taking the ratio,
we obtain $\rho_1 = \cdots = \rho_7 = 0.116$ and 
$\rho_8 = \rho_9 = \rho_{10} = 0.064$,
which are in very good agreement with (\ref{extent_CLM}).



\section{Summary and discussions}
\label{sec:summary}

In this paper we have applied the CLM
to the Euclidean IKKT matrix
model, which is conjectured to be a nonperturbative formulation of
superstring theory in ten dimensions. 
This is the first time that a first principle study 
of this model produced
clear results on the question of 
dynamical compactification of extra dimensions via SSB 
of the SO($10$) rotational symmetry of the model.
Monte Carlo simulations are plagued by a serious sign
problem, which is overcome by applying the CLM.
In order to avoid 
the singular-drift problem in the CLM,
we deform the model by adding a mass-like term in the fermionic action
parameterized by the parameter $\mf$ so that the
conditions proposed in Ref.~\cite{Nagata:2016vkn} can be met. 
By taking the large-$N$ limit and then the $\epsilonm\to 0$ limit, 
we have studied the SSB pattern as a function of the $\mf$. 
At $\mf=3.0$ an SO($7$) vacuum is found, which is the maximally symmetric
vacuum of the deformed theory. At $\mf=1.4$ we find that 
our results are consistent with an SO($4$) vacuum.
At $\mf=1.0$, $0.9$ and $0.7$, the vacuum becomes SO($3$) symmetric.
Taking into account the argument that an SO($2$) vacuum is
not likely to be realized in this 
model \cite{Nishimura:2000ds,Nishimura:2000wf},
we conclude that 
our results are consistent
with the ones obtained by using the GEM on the Euclidean IKKT matrix
model \cite{Nishimura:2011xy}, which predict SSB to an SO($3$) vacuum.

As a consistency check,
we performed independent calculations
applying the GEM to the
$\mf$-deformed Euclidean IKKT matrix model.
We did a
three-loop calculation using SO(7) and SO(6)
ansatzes and calculated the free energy.
It turns out that the transition 
from an SO(7) symmetric vacuum
to an SO(6) symmetric one
occurs smoothly as $\mf$ decreases.
Also the extent of space obtained at $\mf = 3.0$ agrees
very well between the two methods.

Our conclusion that the SO(10) rotational symmetry
of the Euclidean IKKT model breaks down to SO(3) 
due to the phase of the Pfaffian
is interesting, but it makes the model somewhat 
more difficult to interpret.
Given the promising properties of the Lorentzian model 
reviewed in Section \ref{sec_IKKT},
we consider that the naive Wick rotation to the Euclidean model
is not the right direction to pursue.
On the other hand, the fact that the CLM enabled
us to obtain a clear SSB pattern for the deformed model, 
which suffers from a severe sign problem, is encouraging.
We hope that the CLM is equally useful in investigating
the Lorentzian IKKT model, in particular in the presence of
fermionic matrices, which are not included yet in 
Ref.~\cite{Nishimura:2019qal}.

Let us emphasize that the model has the amazing properties that spacetime, 
and possibly the matter content as well, are contained in the matrix
degrees of freedom and that many interesting related questions can be
answered dynamically.
The surmounting evidence that the IKKT matrix
model has nontrivial dynamics
makes it a particularly promising candidate
for a nonperturbative definition of superstring theory.
By improving the algorithms that solve the sign problem and by
acquiring more computational power that will allow us to study
larger $N$, there is a great hope that we will find answers to many
profound questions, in a similar way that it was done for 
QCD using the lattice gauge theory.




\section*{Acknowledgements}

The authors would like to thank S.~Iso, H.~Kawai, H.~Steinacker 
and A.~Tsuchiya for valuable discussions.
T.~A.\ was supported in part
by Grant-in-Aid for Scientific Research (No.\ 17K05425) from Japan
Society for the Promotion of Science.
Computations were carried out
using computational facilities at KEKCC and the NTUA het cluster.
This work was also supported by computational time granted by the
Greek Research \& Technology Network (GRNET) in the National HPC
facility - ARIS - under project ID IKKT10D.

\appendix

\section{The singular-drift problem vanishing at large $N$}
\label{sec:large-N-singular-drift}

\begin{figure}[t]
\centering 
 \includegraphics[width=0.45\textwidth]{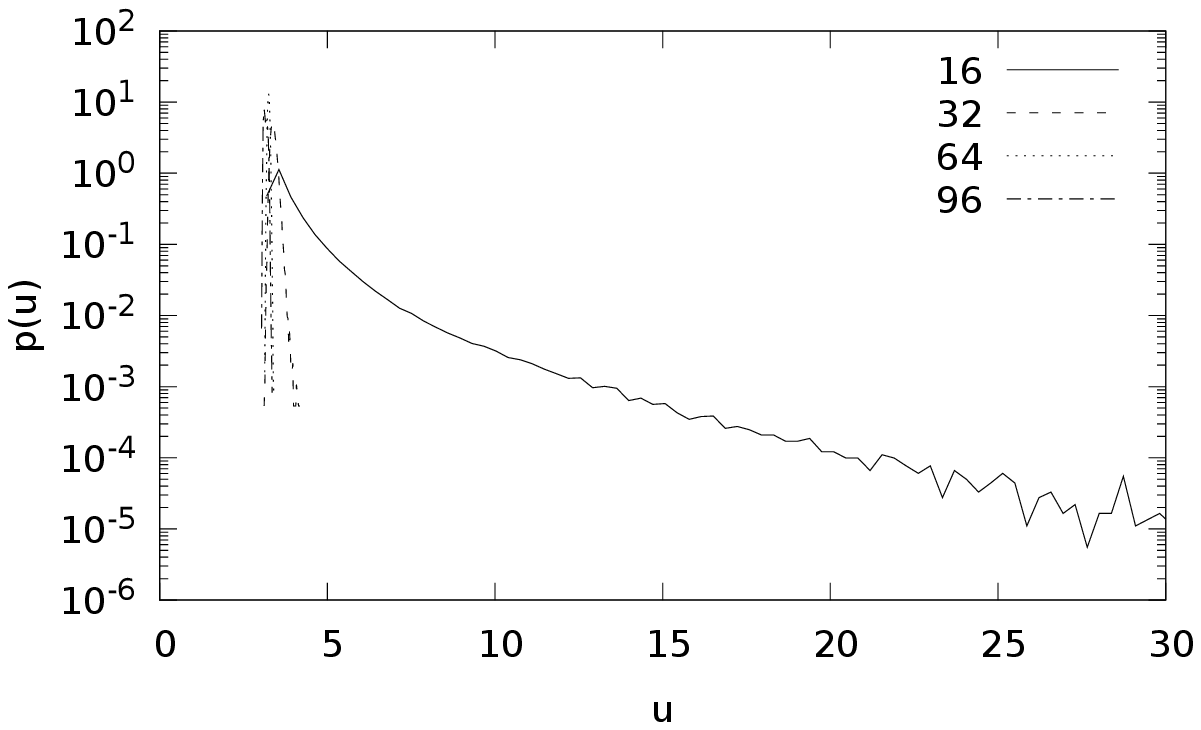}
 \includegraphics[width=0.45\textwidth]{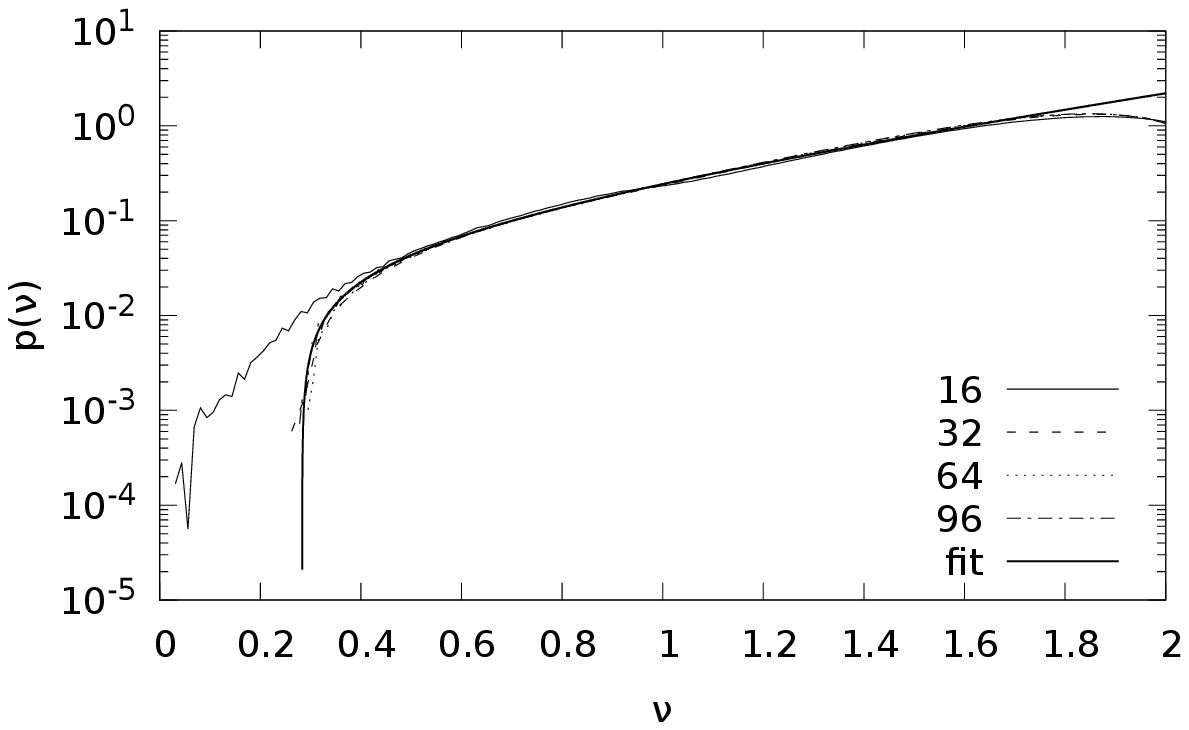}
    \caption{(Left) The distribution
$p(u)$ of $u$ defined in Eq. \protect\rf{m:9} 
for
$m_\mu=$ $(0.5, 0.5,$ $1, 2, 4,$ $8, 8, 8,$ $8, 8)$, 
$\mf=0.9$, $\epsilonm=0.16$. 
For $N=16$ we observe a long subexponential tail indicating the presence
of the singular-drift problem. For $N\geq 32$, where the
singular-drift problem is absent, the tail falls off
(super)exponentially.
(Right) The small $\nu_k$ distribution of the Youla decomposition
values $\nu_k\geq 0$, $k=1,\ldots,n$, $n=8(N^2-1)$ of 
the antisymmetric matrix $\cal M$ for the same parameters,
together with a fit
of the form $(\nu-\nu_{\rm c})^a {\rm e}^{b \nu}$ with $\nu_{\rm c}=0.283(1)$,
$a=0.82(2)$ and $b=1.49(3)$. The cutoff $\nu_{\rm c}$ develops for $N\geq 32$, 
whereas for $N=16$ the values accumulate near zero,
causing the singular-drift problem. 
\label{f:m01} }
\end{figure}

Rather surprisingly, we find
that the singular-drift problem vanishes for large enough $N$
for given values of $(\mf,\epsilonm)$.
In Fig.~\ref{f:m01} (Left), we plot
the probability distribution $p(u)$ of $u$ defined in \rf{m:9} for 
$\mf=0.9$ and $\epsilonm=0.16$ 
with $m_\mu=(0.5, 0.5, 1, 2, 4, 8, 8, 8, 8, 8)$
using various values of $N$ within $16\leq N \leq 96$. 
We observe for $N=16$ that the tail of the $p(u)$ distribution 
is subexponential as a result of the singular-drift 
problem\cite{Nagata:2016vkn}. For $N\geq 32$, the tail of the
distribution is suppressed as an exponential or faster,
indicating that the singular-drift problem is absent. 

This is confirmed 
by measuring the distribution $p(\nu)$ of the fourth root
of the doubly degenerate non-negative eigenvalues $d_k$ 
of the positive definite Hermitian matrix 
${\cal D}={\cal M} {\cal M}^\dagger$, where $\nu_k=d_k^{1/4}$. 
These can be seen to be directly
related to the singular-drift problem by considering the Youla
decomposition \cite{Yula:1961je}
of the antisymmetric\footnote{Note that the antisymmetry of $\cal M$
remains after the addition of the term $\Delta S_{\rm f}$.} even
dimensional $16(N^2-1)\times 16(N^2-1)$ matrix $\cal M$
\begin{equation}
\label{m:12}
{\cal U}^\intercal {\cal M} {\cal U} =
\mbox{diag}\left\{
\begin{pmatrix}0 & \nu_1\\ -\nu_1 & 0\end{pmatrix}
\begin{pmatrix}0 & \nu_2\\ -\nu_2 & 0\end{pmatrix}
\ldots
\begin{pmatrix}0 & \nu_n\\ -\nu_n & 0\end{pmatrix}
\right\}\, ,
\end{equation}
where $\cal U$ is unitary, ${\cal U}^\intercal$ is its transpose
and $\nu_k\geq 0$, $k=1,\ldots,n$, $n=8(N^2-1)$. Then,
$$
\Pf\,{\cal M} = {\rm e}^{i\theta} \prod_{k=1}^n \nu_k
            = {\rm e}^{i\theta}  |\det{\cal M}|^{1/2} 
            = {\rm e}^{i\theta} \det{\cal D}^{1/4} \  ,
$$
where $\det {\cal U}={\rm e}^{i\theta}$.
The singularities of the drift term \rf{m:3} appear from the term
\begin{equation}
\label{m:13}
\frac{1}{2}\Tr\left(
 \frac{\partial {\cal M}}{\partial \left(A_\mu\right)_{ji}}
 {\cal M}^{-1}
\right) = 
i \frac{\partial \theta}{\partial \left(A_\mu\right)_{ji}}
+
\sum_{k=1}^N \frac{1}{\nu_k}
\frac{\partial \nu_k}{\partial \left(A_\mu\right)_{ji}} \  ,
\end{equation}
when the $\nu_k$ accumulate near the origin.
The plot in Fig.~\ref{f:m01} (Right)
shows that the distribution of the
${\nu_k}$ develops a small-$\nu$ finite cutoff $\nu_{\rm c}>0$ for $N\geq 32$
proving that there is no singular-drift problem and that 
for $N=16$, the $\nu_k$ accumulate
near the origin showing the appearance of the singular-drift problem.

\bibliographystyle{JHEP}
\bibliography{ref}



\end{document}